\documentclass[preprint,12pt]{elsarticle}

\usepackage[x11names]{xcolor}
\usepackage{array}
\usepackage{tabularx}
\usepackage{hyperref}
\usepackage{amsfonts} 
\usepackage{breqn}
\usepackage{multirow}
\usepackage{pdflscape}
\usepackage{float}
\usepackage{caption}
\usepackage{float}
\usepackage{bm}
\usepackage{booktabs} 

\usepackage{amssymb}
\usepackage{amsmath}
\usepackage{mathtools}
\usepackage{xcolor}
\usepackage{subfigure}
\usepackage{hyperref}
\usepackage{float} % Add this line to use the float package

\usepackage{bm}   % for \bm macro

\usepackage{adjustbox} % for nice table zwang
\usepackage{multirow} % for nice table zwang
\usepackage{subcaption} % for subfigure zwang

\usepackage[ruled,vlined]{algorithm2e}
\hypersetup{
 colorlinks,
 citecolor=black,
 filecolor=black,
 linkcolor=black,
 urlcolor=black
}
\usepackage{amssymb}
\journal{arXiv}

\biboptions{sort&compress}

\begin{document}

\begin{frontmatter}


%%       addressline={},
%%       city={},
%%       postcode={},
%%       state={},
%%       country={}}

\title{Inferring turbulent velocity and temperature fields and their statistics from Lagrangian velocity measurements using physics-informed Kolmogorov-Arnold Networks}

%%       addressline={},
%%       city={},
%%       postcode={},
%%       state={},
%%       country={}}
%%       addressline={},
%%       city={},
%%       postcode={},
%%       state={},
%%       country={}}

\author[inst1,label1]{Juan Diego Toscano}

\author[inst2, label1]{Theo K\"aufer}
\author[inst3]{Zhibo Wang}
\author[inst1,inst3]{\\ Martin Maxey}
\author[inst2,label2]{ Christian Cierpka}
\author[inst1,inst3,label3]{George Em Karniadakis}
\affiliation[inst1]{organization={Division of Applied Mathematics, Brown University},%Department and Organization
%      addressline={}, 
      city={Providence},
      postcode={02912}, 
      state={RI},
      country={USA}}
\affiliation[inst2]{organization={Institute of Thermodynamics and Fluid Mechanics, Technische Universität Ilmenau},%Department and Organization
      city={Ilmenau},
      postcode={98693}, 
      state={Ilmenau},
      country={Germany}}
\affiliation[inst3]{organization={School of Engineering, Brown University},%Department and Organization
%      addressline={}, 
      city={Providence},
      postcode={02912}, 
      state={RI},
      country={USA}}

\fntext[label1]{The first two authors contributed equally to this work}
\fntext[label2]{Correspondig author: christian.cierpka@tu-ilmenau.de}
\fntext[label3]{Correspondig author: george\_karniadakis@brown.edu}
\begin{abstract}
We propose the Artificial Intelligence Velocimetry-Thermometry (AIVT) method to infer hidden temperature fields from experimental turbulent velocity data. This physics-informed machine learning method enables us to infer continuous temperature fields using only sparse velocity data, hence eliminating the need for direct temperature measurements. Specifically, AIVT is based on physics-informed Kolmogorov-Arnold Networks (not neural networks) and is trained by optimizing a combined loss function that minimizes the residuals of the velocity data, boundary conditions, and the governing equations. We apply AIVT to a unique set of experimental volumetric and simultaneous temperature and velocity data of Rayleigh-Bénard convection (RBC) that we acquired by combining Particle Image Thermometry and Lagrangian Particle Tracking. This allows us to compare AIVT predictions and measurements directly. We demonstrate that we can reconstruct and infer continuous and instantaneous velocity and temperature fields from sparse experimental data at a fidelity comparable to direct numerical simulations (DNS) of turbulence. This, in turn, enables us to compute important quantities for quantifying turbulence, such as fluctuations, viscous and thermal dissipation, and QR distribution. This paradigm shift in processing experimental data using AIVT to infer turbulent fields at DNS-level fidelity is a promising avenue in breaking the current deadlock of quantitative understanding of turbulence at high Reynolds numbers, where DNS is computationally infeasible.
\end{abstract}


\begin{keyword}
turbulence\sep data assimilation \sep physics-informed machine learning \sep experimental methods \sep Kolmogorov-Arnold networks.

\end{keyword}

\end{frontmatter}



\section{Introduction}
\label{Introduction}
Turbulence is considered as the last frontier in classical physics. At terrestrial and astrophysical scales, turbulence is predominantly thermally driven~\cite{Guervilly2019,mapes1993cloud,Schumacher.2020Colloquium,Young.2017Forward}, and hence is characterized by the joint transport of heat, mass and momentum. Besides its occurrence in nature, thermal convection is also prevalent in many engineering and technological applications~\cite{bitharas_interplay_2022,otto_unsteady_2023,st_martin_wind_2016}.
Thus, a sound and detailed knowledge of the underlying mechanism is not only important for a better understanding of our environment but also essential for engineering applications, e.g., a successful transition from fossil fuels to renewable energy~\cite{milan_turbulent_2013}.

Since the pioneering work by Henri B\'enard~\cite{Benard1901} and Lord Rayleigh~\cite{Rayleigh1916}, thermal convection has been studied extensively by various computational and experimental methods alike~\cite{Manneville2006,Ahlers.2009Heat,chilla_new_2012}. While increasingly powerful computers significantly advanced the potential to study thermal turbulence and turbulence flows in general by means of Direct Numerical Simulation (DNS)~\cite{iyer_classical_2020,jansson_exploring_2023}, the immense computational cost constrains their application  to only research problems, simple geometries, and moderate Reynolds numbers.

In contrast to simulations where assumptions and simplifications are necessary, experimental methods capture the physics inherently. While flow measurement techniques have advanced enormously in the last decades~\cite{Raffel.2018Particle,Kahler_2016}, they often have a limited spatiotemporal resolution, cover only a limited region, and have their method-specific uncertainties. Furthermore, measuring different quantities simultaneously, e.g., temperature and velocity in thermal convection, drastically increases the complexity.
Therefore, only a few cases of volumetric combined temperature and velocity measurements have been reported~\cite{deng2022combined,massing2018volumetric,segura2015simultaneous,stelter2023thermographic,rietz2014combined,Schiepel.2021Simultaneous,kaufer_volumetric_2024}. Out of these studies only Käufer and Cierpka~\cite{kaufer_volumetric_2024} provide a unique data set of spatiotemporal resolved joint Lagrangian measurements of velocity and temperature in turbulent convection in a macroscopic setup.

Recently, physics-informed neural networks (PINNs) have arisen as an alternative for studying thermal convection~\cite{kag2022physics,du2023state} and turbulence \cite{pioch2023turbulence,patel2024turbulence}.
The PINN approach uses a multilayer perceptron (MLP) to approximate the solution of ordinary/partial differential equations (ODE/PDE) by minimizing a multi-objective loss function that attempts to fit any observable data while satisfying the underlying physical laws~\cite{raissi2019physics,raissi2020hidden,boster2023artificial}. PINNs extensions, such as hidden fluid mechanics (HFM) ~\cite{raissi2020hidden} or artificial intelligence velocimetry (AIV) \cite{cai2021physics,boster2023artificial}, have shown remarkable performance in inverse problems in fluid dynamics and beyond. These methods exploit PINNs' flexibility to obtain solutions with partial or no knowledge of the boundary conditions from only sparse measurements of unsteady velocity fields.
However, PINNs training involves minimizing a multi-objective loss function, which, if not appropriately optimized, induces imbalances leading to partial enforcement of the desired constraints, especially when dealing with highly nonlinear, multi-scale, or chaotic behavior problems~\cite{anagnostopoulos2024residual}. Thus, most studies in thermal convection and turbulence so far with PINNs have been  limited to synthetic data and idealized simulations.

To address these issues, researchers have proposed several methods to improve the PINNs performance that can be roughly classified into three classes, namely, representation model modifications~\cite{wang2021eigenvector,sukumar2022exact,salimans2016weight,wang2021understanding,wang2024piratenets,zhang2023artificial,howard2024finite}, PDE reformulations ~\cite{jin2021nsfnets,wang2023solution,basir2022investigating} or optimization algorithm enhancements~\cite{anagnostopoulos2024residual,wang2021understanding,wang2024respecting,mcclenny2023self,lu2021deepxde,wu2023comprehensive,chen2024self}. Several studies have found that the optimal frameworks are obtained by combining some of this methods. For instance, Shukla et al.~\cite{shukla2024comprehensive} introduced Chebyshev physics-informed Kolmogorov-Arnold networks (cPIKANs). By replacing the MLP with a modified Kolmogorov-Arnold Network~\cite{liu2024kan}, the authors showed that cPIKANs, combined with other enhancements such residual-based attention (RBA)~\cite{anagnostopoulos2024residual}, require fewer parameters, are more robust to noise, and can potentially outperform PINNs. These methods extend the physics-informed machine learning framework to solve challenging problems, possibly making them suitable candidates for turbulent thermal convection problems. 

Combining reliable experimental data on thermal convection with scientific machine learning can significantly benefit fluid mechanics research by leveraging the advantages of both measurements and scientific modeling while overcoming their individual limitations.

\section{Problem Description} 

\begin{figure}[H]
\centering
\includegraphics[width=\textwidth]{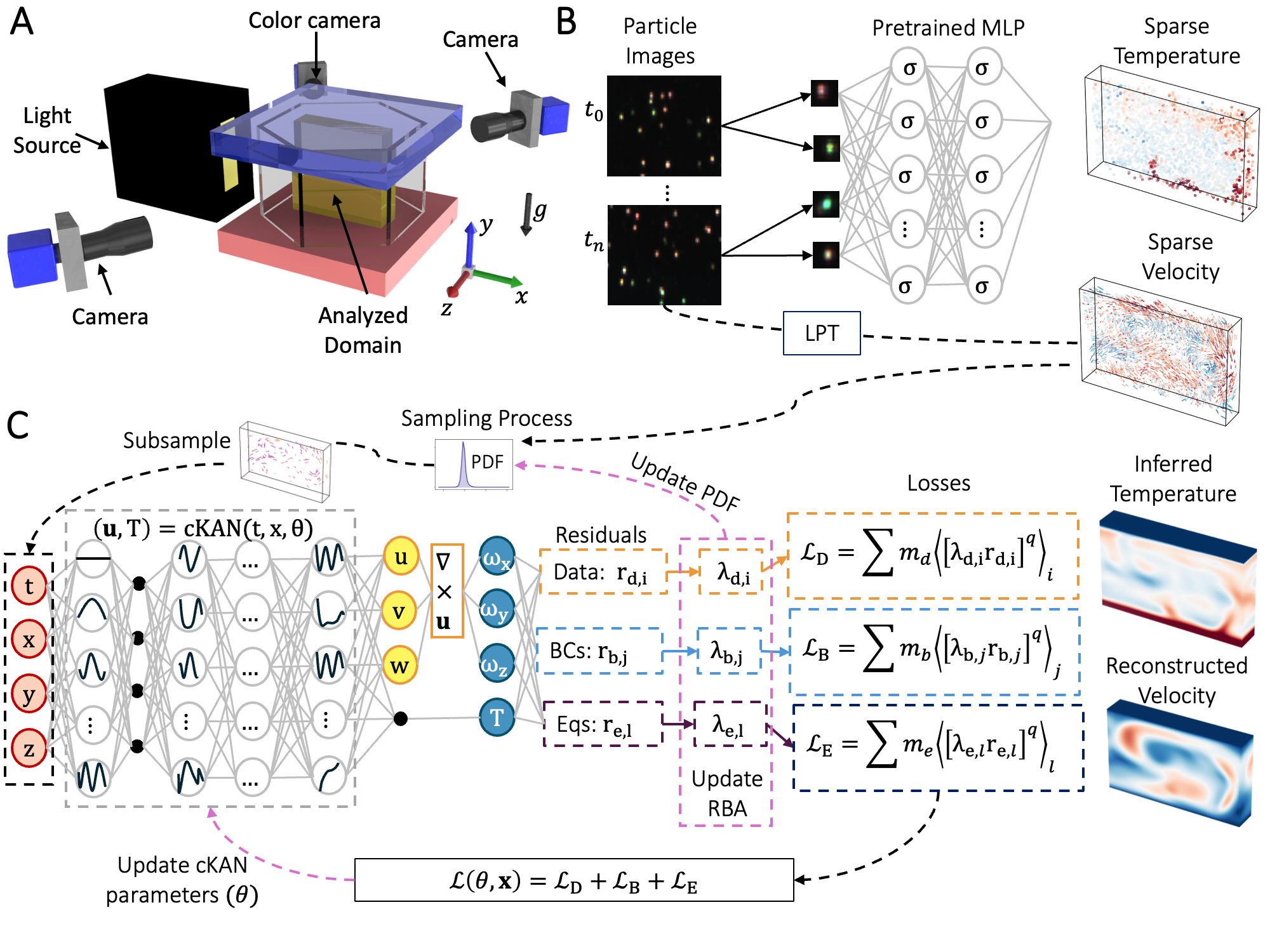}
\caption{\textbf{Problem Setup}. (\textbf{A}) The experimental setup includes a hexagonal RBC cell, cameras, and a light source. The illuminated vertical slice is indicated in yellow. Gravity $g$ (black arrow) acts in a negative $y$ direction. The experiment was performed at Rayleigh number Ra = $3.4\times10^7$ and Prandtl number Pr = 10.6. (\textbf{B}) Schematic view of the joint temperature and velocity measurement processing. The particle positions and velocity vectors are obtained using the Shake-the-Box method (DAVIS 10.2, LaVision GmbH) \cite{schanz_shake--box_2016}. The particle temperature is derived from the individual particle color images using a multi-layer perceptron pre-trained on calibration data. (\textbf{C}) At each training iteration, the domain is sampled based on a PDF that identifies high-error regions. These coordinates are fed into a modified Kolmogorov Arnold network (cKAN) that predicts 3D velocities and temperature. Residuals for data, boundary conditions, and equations are calculated, and derivatives for equation constraints are obtained using automatic differentiation. Residual-based attention (RBA) multipliers are updated using an exponentially weighted moving average of the residuals, and a PDF is computed for the next iteration's sampling. Residuals are scaled with RBA, balancing the local contribution of each training point. The mean of scaled residuals to a power $q$ forms each loss subterm. The total loss updates cKAN parameters, resulting in continuous and differentiable temperature and velocity fields. }
\label{fig:VW-KAIV}
\end{figure}

We obtain the experimental data by combining Particle Image Thermometry (PIT) and Lagrangian Particle Tracking (LPT). In particular, we use Thermochromic Liquid Crystals (TLC) as tracer particles to study Rayleigh-B\'enard convection at Ra~=~$3.4\times10^7$ and Pr~=~10.6 in a hexagonal container. As shown in Figure~\ref{fig:VW-KAIV}(A), we studied an extended vertical slice at the cell center using two monochrome and one color camera. The recorded images were processed by LPT, which provides three-dimensional (3D) particle positions and velocities (see Figure~\ref{fig:VW-KAIV} (B)). To determine the temperature information, we feed the individual color particle images into a pre-trained Multi-Layer Perceptron (MLP) that estimates the particle temperature from the corresponding particle's color image. Combining these approaches, we obtain about 1,000 snapshots with approximately 3,000 Lagrangian velocity and temperature data points per snapshot, of which we use 282 snapshots corresponding to 20 non-dimensional times scales. Further details on the setup and the methods can be found in section~\ref{sec:materials_and_methods} and reference~\cite{kaufer_volumetric_2024}.

Based on these experimental data and the flexibility of physics-informed machine learning to incorporate the underlying physical laws \cite{cai2021artificial,boster2023artificial,raissi2020hidden}, we introduce Artificial Intelligence Velocimetry-Thermometry (AIVT). AIVT is a scientific machine learning method based on cPIKANs~\cite{shukla2024comprehensive} and AIV~\cite{cai2021artificial,boster2023artificial} that uses the full set of Navier-Stokes equations including the energy equation to model turbulent thermal convection~\cite{chilla_new_2012}. As shown in Figure~\ref{fig:VW-KAIV} (C), we use AIVT to infer 3D continuous and differentiable velocity ($\bm{u}=(u,v,w)$) and temperature ($T$) fields from the discrete experimental observations of velocity. We train our model by optimizing a combined loss function ($\mathcal{L}$) that minimizes the residuals (i.e., point-wise error) of the velocity data $(r_{d})$, boundary conditions $(r_{b})$, and governing equations $(r_{e})$. 

To deal with the local imbalances related to optimizing $\mathcal{L}$, we introduce a residual-based attention method with resampling (RBA-R). RBA-R uses residual-based attention (RBA)\cite{anagnostopoulos2024residual} weights ($\lambda_i$) as local multipliers to balance the point-wise errors, enabling a uniform convergence along the analyzed domain~\cite{anagnostopoulos2024learning}. Additionally, since $\lambda_i$ contain historical information of the residuals, we employ them to compute an RBA-based probability density function (PDF) used to resample the high error regions. To ensure the exact enforcement of the constraints, we follow Sukumar and Srivastava~\cite{sukumar2022exact} and use approximate distance functions (ADF) to impose the temperature boundary conditions and redesign the base KAN model to infer divergence-free fields. 

Finally, to simplify the inherent optimization problem related to training a multi-objective loss function, we propose a sequential learning approach. In the first step, we solve a purely data-driven problem where the model mainly fits the data and the boundary conditions. Subsequently, we train the model using a lower Rayleigh (Ra) number (i.e., ``partial physics" ), which enables capturing the diffusive features of the desired solution. The first steps can be considered as a ``guided" initialization that prepares the model to learn the desired temperature and velocity fields at higher Ra. The specific details of this method along with the corresponding ablation study are detailed in section~\ref{sec:materials_and_methods} and Table~\ref{ablation}, respectively.

We train our AIVT model with 50\% of the experimental velocity measurements in the core region ($0.1<y<0.9$) and validate our results with the remaining unseen data. Since the outputs of our model are continuous and differentiable, we derive additional 3D fields such as convective heat transfer, vorticity, and viscous and thermal dissipation rates that are hard to measure or not even directly measurable. Additionally, we compute the turbulent flow statistics, including mean fields, root-mean-squared fluctuations, viscous and thermal dissipation rates, and the $Q$ and $R$ invariants of the velocity gradient tensor. All quantities are obtained in dimensionless form based on the characteristic units (see \eqref{eq:u_dim}. We compare the results with the available measurement data and values from the literature and find good agreement.
\section{Methods}
\label{sec:materials_and_methods}
\subsection{Gathering of the experimental temperature and velocity data}
\label{Sec:methods_exp}
To obtain the joint temperature and velocity data by combining Lagrangian Particle Tracking (LPT) and Particle Image Thermometry (PIT), temperature-sensitive encapsulated Thermochromic Liquid Crystals (TLCs) are used to visualize RBC as described in~\cite{kaufer_volumetric_2024}. 
The experiment was performed in an equilateral hexagonal Rayleigh-B\'enard convection cell of height $h$= 60 mm and width of $w$= 104 mm between the parallel sides. The dimension of the convection cell corresponds to an aspect ratio $\Gamma=w/h=1.73$.
The sidewalls are made from glass to allow optical access, and the top and bottom plates are made from aluminum to approximate isothermal boundary conditions.
The flow was illuminated by a white light LED and observed by three cameras, one of which is color-sensitive, as shown in Figure~\ref{fig:VW-KAIV} (A). The camera setup and the illuminated domain resulted in a volume of interest of $\approx 80 \times 60 \times 12$ mm$^3$ ($x \times y \times z$) at the center of the cell. The working fluid was a water-glycerol mixture, 13 $\%$ glycerol by volume, resulting in a Prandtl number Pr=~$\nu/\kappa$=~10.6. For the experiment, a temperature difference of $\Delta T=10.3 ^\circ$C was applied, which resulted in a Rayleigh number Ra=~$\alpha g \Delta T h^3/ \nu \kappa=3.4\times 10^7$. Here $\alpha$, $g$, $\nu$, and $\kappa$ denote the thermal expansion coefficient, the acceleration due to gravity, height, kinematic viscosity, and thermal diffusivity, respectively.

During the experimental run, images were recorded by all cameras at a rate of 10 Hz, and the data was processed as visualized in Fig.~\ref{fig:VW-KAIV} (B).
The particle positions and velocity were obtained from the image data by applying the Shake-the-Box algorithm (DAVIS 10.2, LaVision GmbH). Afterward, the particle position in space was back-projected into the particle image of the color camera. The temperature of the individual particle image was measured by using the $5\times5$ pixel particle color image and the coordinate of the center pixel in the color camera image as an input to a densely connected neural network that was trained on a set of calibration data.
The densely-connected neural network consists of two layers with 10 neurons each and ReLU activation. To obtain the training data, uniform temperature distributions were established in the RBC cell at several reference temperature steps ranging from 19.6~$^\circ$C to 22.6~$^\circ$C in 0.2 $^\circ$C steps. For each reference temperature, the temperature was measured by sensors in the plates, while color images were recorded by the color camera. Subsequently, the individual particle images were extracted and used to train the neural network to predict the temperature with the sensor temperature serving in reference. When applying the trained neural network to a set of unseen test data, standard deviations below 0.25 $^\circ$C were obtained, resulting in relative uncertainties of about 8$\%$.

The raw measured temperature data was post-processed by applying a sliding median filter and discarding tracks that were shorter than seven-time steps,  resulting in about 3000 Lagrangian joint temperature and velocity data per snapshot. For further analysis, the data was non-dimensionalized by their respective characteristic units:
\begin{gather}
t=\frac{t_{\mathrm{dim}}}{\sqrt{h / \alpha g\Delta T}} , \quad
\mathbf{x}=(x, y, z)=\frac{\mathbf{x}_{\mathrm{dim}}}{h},\nonumber \\
T(\mathbf{x}, t)=\frac{T_{\mathrm{dim}}(\mathbf{x}, t)-T_{\mathrm{c,dim}}}{\Delta T_{\mathrm{dim}}},\quad
\mathbf{u}(\mathbf{x}, t)=\frac{\mathbf{u}_{\mathrm{dim}}(\mathbf{x}, t)}{\sqrt{h \alpha g\Delta T_{\mathrm{dim}}}}\label{eq:u_dim}
\end{gather}
In these equations $t_{\mathrm{dim}}$, $\mathbf{x}_{\mathrm{dim}}$ $T_{\mathrm{dim}}$, $\mathbf{u}_{\mathrm{dim}}$, denote the dimension-attached time, length, temperature, and velocity.
Thus, $t$, $\mathbf{x}$ $T$, $\mathbf{u}$ refer to the time, length, temperature, and velocity in dimensionless units.

\subsection{Underlying Physical Laws} We consider the flow in the Rayleigh-Bénard convection cell under the Boussinesq approximation (i.e., full set of Navier-Stokes equation including the energy equation). In this study, we use the velocity-vorticity (VW) formulation, which, given its independence of pressure, allows us to infer temperature directly from sparse velocity observations and boundary conditions. The detailed mathematical formulation is described in ~\ref{Uderlying_Laws}.

The analyzed domain $\bm{x}=(x,y,z)\in\Omega=(-0.6,0.6)\times(0,1)\times(-0.1,0.1)$ is a cuboid located at the center of the Rayleigh-Benard hexagonal cell (See Figure~\ref{fig:VW-KAIV})(A). Notice that gravity acts in the negative $y$ direction.

 \subsection{Definition of the analyzed quantities}
Here, we provide the definition of the quantities analyzed in the main part, which might not be familiar to the reader.
\begin{enumerate}
  \item Temperature fluctuations:
  \begin{equation}
    \Theta=T(\mathbf{x},t)-\langle T \rangle _{V,t}.
    \label{eq:theta}
  \end{equation}
  \noindent with $\langle T \rangle_{V,t}$ denoting the mean temperature over the analyzed domain of the experiment (i.e., volume and time).
  \item Convective heat transfer:
  \begin{equation}
    J = \sqrt{Ra Pr} \thickspace v\left( \mathbf{x}, t\right) \Theta \left( \mathbf{x}, t \right).
    \label{eq:conv_heat_transfer}
\end{equation}
\item Fluctuation profiles:
\begin{align}
\label{u_fluct_222}
  u'(t,\bm{x})&=u(t,\bm{x})-\langle u\rangle_{A,t}\\
\label{v_fluct_222}
  v'(t,\bm{x})&=v(t,\bm{x})-\langle v\rangle_{A,t}\\
\label{w_fluct_222}
  w'(t,\bm{x})&=w(t,\bm{x})-\langle w\rangle_{A,t}\\
\label{T_fluct_222}
  T'(t,\bm{x})&=T(t,\bm{x})-\langle T\rangle_{A,t},
\end{align}
\noindent where $\langle u\rangle_{A,t}$, $\langle v\rangle_{A,t}$, $\langle w\rangle_{A,t}$, $\langle T\rangle_{A,t}$ are the mean velocities and temperature vertical profiles, which are functions of $y$; $A$ is the cross-sectional area and $t$ is the time.

\item Thermal boundary layer thickness:
  \begin{equation}
\delta _T= \frac{1}{2\mathrm{Nu}}
 \label{eq:T_boundary_layer}
\end{equation}
with the Nusselt number Nu= 22 based on Ra and Pr according to figure 7 of ref.~\cite{stevens_unifying_2013}.
\item Viscous boundary layer thickness:
  \begin{equation}
    \delta_\nu=\delta_T \sqrt{\mathrm{Pr}}.
    \label{eq:vicous_boundary_layer}
  \end{equation}
  \item Theoretical temperature profile for half height according to~\cite{shishkina_mean_2009}:
  \begin{equation}
    T^*=1-\exp (-\delta_T-0.5 \delta_T ^2).
  \end{equation}
\item $Q$ and $R$ invariants of the velocity gradient tensor $\mathbf{A}$:
\begin{align}
Q=-\frac{1}{2} A_{i m} A_{m i}\\
R=-\frac{1}{3} A_{i m} A_{m n} A_{n i},
\label{eq:invariants}
\end{align}
\noindent where $A_{i j}=\frac{\partial u_i}{\partial x_j}.$
\item Viscous dissipation rate:
  \begin{equation}
 \epsilon_K=\frac{1}{2} \sqrt{\frac{\mathrm{Pr}}{\mathrm{Ra}}}\left[(\nabla \boldsymbol{u})+(\nabla \boldsymbol{u})^T\right]^2. 
 \label{eq:kin_dissip}
\end{equation}
\item Thermal dissipation rate: 
\begin{equation}
 \epsilon_T=\frac{1}{\sqrt{R a P r}}(\boldsymbol{\nabla} T)^2.
 \label{eq:therm_dissip}
\end{equation}
\end{enumerate}

\subsection{Artificial Intelligence Velocimetry-Thermometry (AIVT)} AIVT is a scientific machine learning model based on PIKANs~\cite{shukla2024comprehensive} and inspired by AIV~\cite{cai2021artificial,boster2023artificial} that can infer and reconstruct temperature and flow fields from experimental data and the underlying physical laws. This formulation approximates the solution of PDEs using a modified Kolmogorov-Arnold Network(cKAN) with Chebyshev polynomials as univariate functions. In particular, a two-layer cKAN can be represented as follows:
\begin{equation}
 cKAN(\bm{\zeta})=\sum_{q=1}^{n_1}\Phi_q\left(\sum_{p=1}^{n_0}\phi_{q,p}(\zeta_p)\right), 
\end{equation}

\noindent where $n_0, n_1$ are the number of neurons per layer, $\zeta_p$ is a one-dimensional input, and $\phi_{q,p}:[0,1]\rightarrow\mathbb{R}$, $\Phi_q:\mathbb{R}\rightarrow \mathbb{R}$ are learnable univariate functions. 

We use AIVT to obtain continuous and differentiable flow and temperature fields from sparse velocity measurements. In particular, we approximate the solutions of the governing equations as follows:
\begin{equation}
  \label{KAN_vw}
  (\bm{u}, T') = cKAN(t, \textbf{x}, \theta)
\end{equation}
where $\textbf{x} = (x, y, z)$ represents the inputs, with $x, y, z$ as the spatial nondimensional coordinates, $t$ as time. $\bm{u} = (u, v, w)$ is the velocity field used to derive the vorticity vector $\bm{\omega} = (\omega_x,\omega_y,\omega_z)$ by using automatic differentiation (i.e., $\bm{\omega} = \nabla \times \bm{u}$). The temperature $T$ is obtained from the predicted temperature fluctuation $T'$ using approximate distance functions \cite{sukumar2022exact}, which exactly satisfies the temperature boundary conditions.

We impose the remaining constraints by optimizing a combined loss function that minimizes the error from data, boundary conditions, and equations. Each loss term can be represented as:
\begin{align}
\label{data_eq}\mathcal{L}_C(X_C,\theta)&=\sum_{\alpha} m_\alpha\langle[\lambda_{\alpha,i}r_\alpha(\textbf{x}_i,\theta)]^q\rangle_i\text{, where } \textbf{x}_i\in\Omega_C
\end{align}
\noindent where $C=\{D, B, E\}$ is an index that specifies the specific loss group, namely, data ($\mathcal{L}_D$), boundary ($\mathcal{L}_B$) and PDE ($\mathcal{L}_E$). $\langle\cdot\rangle_i$ is the mean operator of the training points $\textbf{x}_i=(t_i, x_i, y_i, z_i)$ in the subset $X_C$ sampled iteratively with a probability function $p_{C,\alpha}$ from the domain $\Omega_C$. Here, $q$ is a positive exponent that controls the loss function's smoothness, enabling us to switch from $L^2$ to $L^1$ norm while training. Each loss group $C$ controls different variables or quantities which are identified by the index $\alpha$; for instance, for data loss (i.e., $C=D$), we constrain the velocity components so, $\alpha=\{u,v,w\}$. The details of the remaining loss terms are specified in ~\ref{AIVT_decription}. The residual $r_\alpha(\textbf{x}_i, \theta) = |\hat{\alpha}(\textbf{x}_i) - \alpha(\textbf{x}_i, \theta)|$ quantifies the mismatch between the prediction $\alpha(\textbf{x}_i, \theta)$ and the ideal value $\hat{\alpha}(\textbf{x}_i)$ (e.g., experimental observation) at point $\textbf{x}_i \in \Omega_C$. We use local weights ($\lambda_{\alpha, i}$) to balance the point-wise contribution of the residual $r_\alpha(\textbf{x}_i, \theta)$ and global weights ($m_\alpha$) to scale the averaged value of subcomponent $\alpha$.

\subsubsection{Sequential Training} Training AIVT involves minimizing 15 objective functions, respectively, which significantly complicates the optimization process.
To simplify this problem, we propose a sequential learning approach that divides the training into four stages. In the first phase, we solve a purely data-driven problem where the models learn the velocity data, boundary conditions, and an initial guess for the temperature (i.e., $T=0.5$ in the core region $0.1<y<0.9$). This initial guess is imposed weakly through a loss function, and its contribution quickly decays in the next stages, enabling the model to learn the actual temperature distribution. In the second step, we partially introduce the governing equation knowledge by optimizing a loss function that enforces a ``simpler" physical law, which helps the model identify the main features of the flow. The first two stages can be considered as a guided initialization where the model learns a similar solution, facilitating and enabling convergence to the actual flow and temperature fields, which are learned during phase three. Once the model learned the desired solution, we set $q=1$ (i.e., switch to $L^1$ norm), which helps refine the details of the learned flow field. The specific details of this approach can be found in~\ref{AIVT_decription}.

\subsubsection{Residual-Based Attention with resampling (RBA-R)} One of the main challenges in training neural networks is that the residuals (i.e., point-wise errors) may get overlooked when calculating a cumulative loss function \cite{mcclenny2023self,anagnostopoulos2024learning}. To address this issue, we employ RBA \cite{anagnostopoulos2024residual} as local weights($\lambda_{\alpha, i}$), which help balance specific residuals' point-wise contribution within each loss term $\alpha$. The update rule for an RBA weight ($\lambda_{\alpha, i}$) for the loss term $\alpha$ and point $x_i$ is based on the exponentially weighted moving average of the residuals defined as:

 \begin{equation}
 \lambda_{\alpha,i}^{(k+1)} \leftarrow (1-\eta)\lambda_{\alpha,i}^{(k)}+\eta {\frac{r_{\alpha,i}^{(k)}}{\;\,\,\lVert \bm{r}_\alpha^{(k)} \rVert_{\infty}}}, \ \ i \in \{0, 1, ..., N\},
 \label{Update_RBA}
 \end{equation}
   
  \noindent where $k$ is the iteration, $N$ is the number of training points, $r_{\alpha, i}$, is the residual of loss term $\alpha$ for point $i$, $\eta$ is a learning rate, and $\gamma$ is a decay term that reduces the contribution of the previous iterations. This formulation induces RBA to work as an attention mask that helps the optimizer focus on capturing the spatial or temporal characteristics of the specific problem \cite{anagnostopoulos2024residual,anagnostopoulos2024learning}. However, RBAs are updated based on the residuals calculated in a specific iteration, which may cause problems for large datasets that require batch training. This residual dependency may induce a mismatch between the network state and the generated attention mask. To address this issue, we propose using the obtained multipliers to resample the critical points. As shown in equation~\ref{Update_RBA}, RBA weights contain historical information about the high-error regions, making them suitable for defining a probability density function $p_{\alpha}(\textbf{x})$. Building on the previous studies \cite{lu2021deepxde,wu2023comprehensive}, we define $p(\textbf{x})^{(k)}$ at iteration $k$ as follows:
   \begin{equation}
   p_{\alpha}^{(k+1)}(\textbf{x})=\frac{(\bm{\lambda}_{\alpha}^{(k)})^{\nu}}{\mathbb{E}[(\bm{\lambda}_{\alpha}^{(k)})^{\nu}]}+c
 \label{Update_pdf}
 \end{equation} 
 \noindent where $(\bm{\lambda}_{\alpha}^{(k)})^{\nu}=\{\lambda_0^{(k)\nu},\lambda_{\alpha,1}^{(k)\nu},...,\lambda_{\alpha,N}^{(k)\nu}\}$ are the RBA weights of loss term $\alpha$. 
 
 The exponent $\nu$ is an integer that controls the standard deviation of $p^k(\textbf{x})$, and $c>0$ is a scalar that ensures that all points are eventually resampled. The main difference with previous approaches \cite{lu2021deepxde,wu2023comprehensive} is that the PDF is based on $\lambda_i$ instead of $r_i$. Since $\lambda_i$ is computed iteration-wise using the cumulative residuals (i.e., historical data), its corresponding PDF is more stable, which enables us to sample $\bm{x}$ at every iteration with negligible computational cost.
\section{Results}

\subsection{Reconstructed velocity} 

To assess the capability of the AIVT method to reconstruct the velocity field, we compare the model predictions on the validation dataset (i.e., 50\% of the measured velocity data). The average relative $L^2$ error on the core region ($0.1<y<0.9$), for $u$, $v$, and $w$ are 9.6\%, 10.8\% and 11.9\%, respectively, and the uncertainty due to the model parameters is less than 0.25\%. The details per velocity component are shown in Table~\ref{Results_apendix}. 

\begin{figure}[H]
\centering
\includegraphics[width=0.6\textwidth]{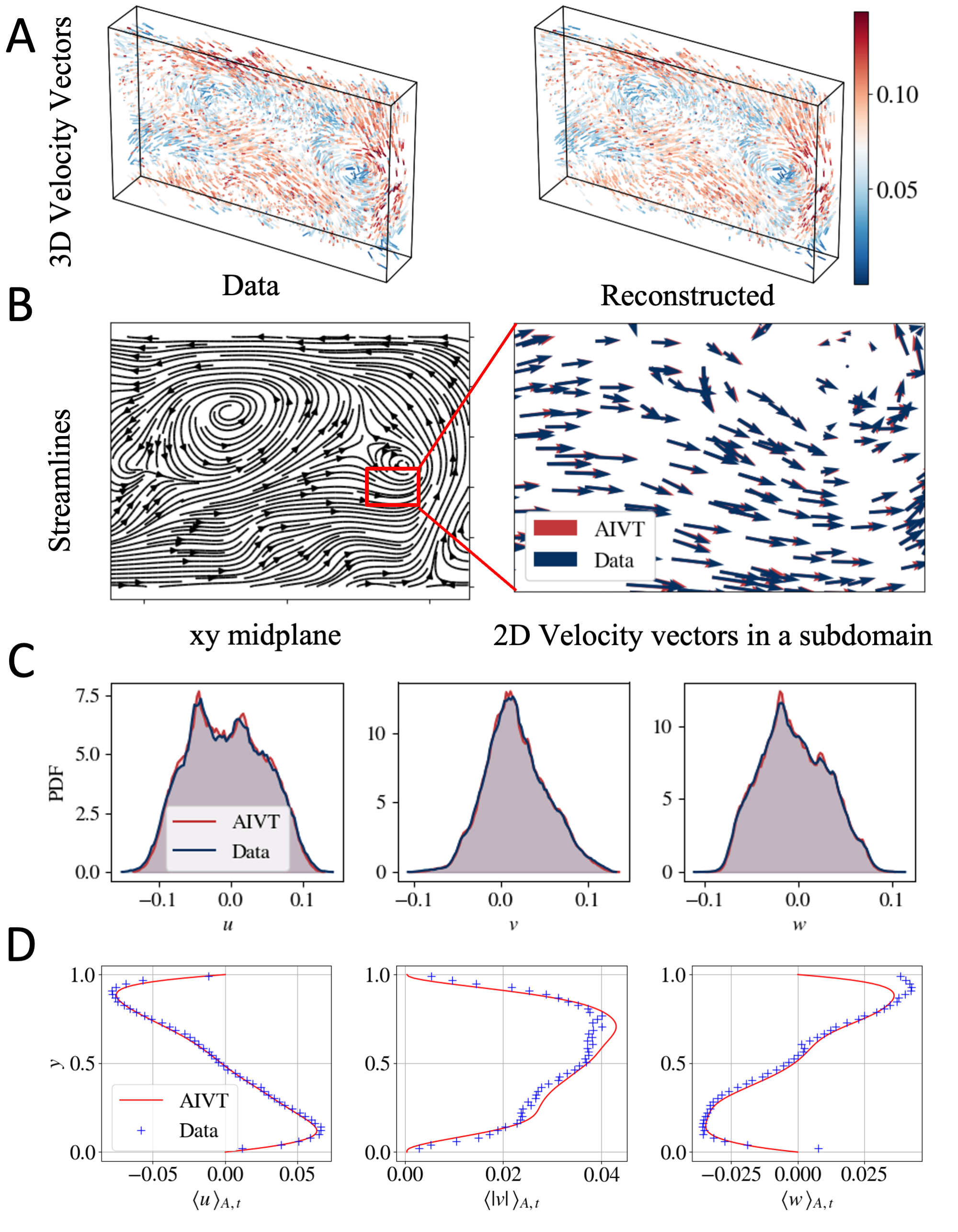}
\caption{\textbf{Velocity reconstruction}. (\textbf{A}) Exemplary instantaneous velocity vectors of the measured and reconstructed velocity at the particle positions in 3D space. The velocity magnitude is color-coded. (\textbf{B}) Streamline plot of the $u$ and $v$ velocity in the $x-y$ plane at $z=0$ and detailed view of the measured (blue) and inferred velocity (red).(\textbf{C}) PDFs of the measured (blue) and reconstructed (red) velocity components at particle locations. (\textbf{D}) Vertical profile of mean velocity fields in the $x$ ($\langle u \rangle _{A,t}$), $y$ ($\langle |v| \rangle _{A,t}$) and $z$ ($\langle w \rangle _{A,t}$) directions. Based on the profiles of horizontal velocity components, the orientation and rotation direction of the LSC can be determined.}
\label{fig:velocity_results}
\end{figure}

Figure~\ref{fig:velocity_results}(A) shows the 3D vector plot of the measured and reconstructed velocities at particle locations for an exemplary snapshot. When comparing the plots, we see that measured and reconstructed velocities are almost indistinguishable. Figure~\ref{fig:velocity_results}(B) shows the streamlines from $x$ and $y$ velocity components in the $x-y$ plane. The streamline plot indicates the presence of the well-known large-scale circulation (LSC) as it is typical for thermal convection with aspect ratios close to unity~\cite{brown2005reorientation}. To analyze the difference between the prediction and measured data, we show a detailed comparison of the measured (blue) and the reconstructed (red) velocity vectors in a 2D subdomain. The detailed view of the velocity vectors confirms that the measured and reconstructed velocities match closely. To obtain a global measure of the reconstruction quality, we compare the PDFs (see Figure~\ref{fig:velocity_results}(C)) of the three velocity components obtained from the measurements (blue) and reconstruction (red). For all velocity components, we observe that PDFs of the measurements and reconstructions align.
\label{Results}

To compare the measured and reconstructed mean velocity fields, Figure~\ref{fig:velocity_results}(D) shows the averaged vertical velocity profiles $\langle u\,\rangle _{A,t}$, $\langle |v|\,\rangle _{A,t}$ and $\langle w\,\rangle _{A,t}$ (red) and the respective binned and averaged measured velocity components (blue crosses). Note that, unlike the PDFs of the velocity, we compare the high-resolution reconstructed velocity with the sparse binned velocity measurements.
Comparing the profiles, we observe that the reconstructed and measured $\langle u\,\rangle _{A,t}$ profiles collapse. Both profiles show the highest and lowest velocity at $y\approx0.1$ and $y\approx0.9$, respectively, with zero or close to zero mean $\langle u\,\rangle _{A,t}$ at $y=\{0,0.5,1\}$. The shape of the profile is representative of the LSC.
Also, the profiles of $\langle |v|\,\rangle _{A,t}$ align; both profiles indicate the highest velocity magnitude at $y \approx 0.7$. The limited volume of investigation only partially captures the LSC, which causes the asymmetry of the profiles with respect to $y=0.5$. The profiles of $\langle w\,\rangle _{A,t}$ show a similar shape to those of $\langle u\,\rangle _{A,t}$ albeit with a flipped sign and at half the magnitude. From the combined information of the $\langle u\,\rangle _{A,t}$ and $\langle w\,\rangle _{A,t}$ profiles, we can determine the orientation and rotation direction of the LSC.
Comparing the measured and reconstructed profiles, we observe that both profiles coincide with the region close to the top plate. In this region, $\langle w\,\rangle _{A,t}$ is overestimated by the measurements. Due to the limited depth of the domain and the camera positioning, the determination of the $z$ position is the most difficult one. Furthermore, particles close to the cooling plate are far below the sensitivity range of the TLCs and thus appear almost transparent. The comparison, however, shows that the AIVT model can successfully reconstruct velocities in regions hard to access by measurements (typically close to the walls) and validates our previous observation of closely matching reconstruction results statistically. 
\begin{figure}[H]%[tbhp]
\centering
\includegraphics[width=0.6\textwidth]{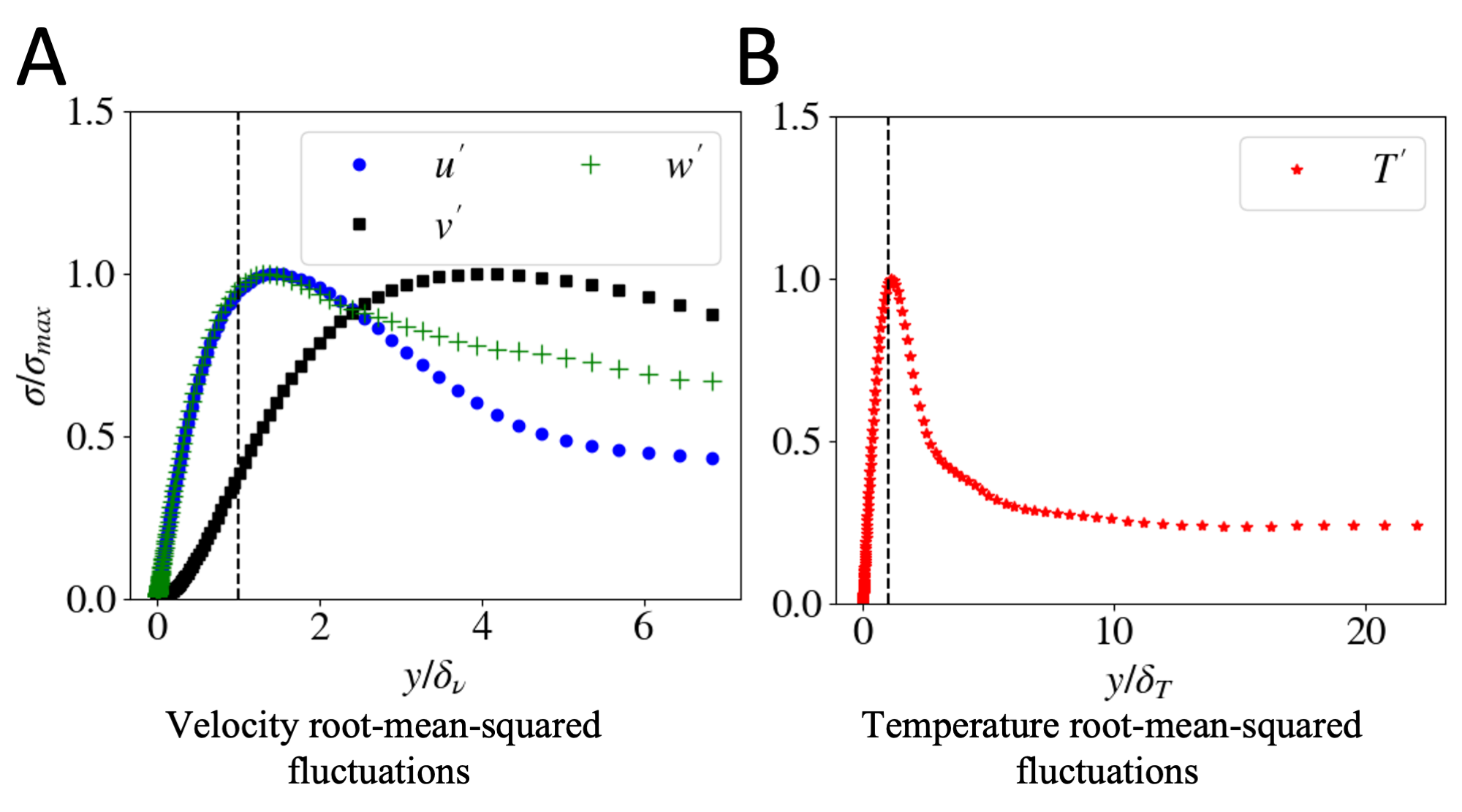}
\caption{\textbf{Velocity and temperature fluctuations profiles.} (\textbf{A}) Normalized root-mean-squared  
($\sigma/(\sigma)_{\max}$) velocity $u^\prime, v^\prime, w^\prime$ and (\textbf{B}) temperature $T^\prime$ fluctuations on the upper half domain. The vertical dashed lines indicate the viscous $\delta_{\nu}$ and thermal $\delta_T$ boundary layer thicknesses based on equation~\eqref{eq:vicous_boundary_layer} and equation~\eqref{eq:T_boundary_layer}, respectively. The profiles are consistent with results reported in~\cite{zhang_statistics_2017,lui_spatial_1998,zhou_thermal_2013} and scaling theory~\cite{grossmann_scaling_2000,stevens2013unifying}.}
\label{fig:rms_R}
\end{figure}

To further investigate the velocity reconstruction performance of the AIVT model in the top boundary region, we show the normalized root-mean-squared 
($\sigma/(\sigma)_{\max}$) profiles of the velocity fluctuations, defined in equations~\ref{u_fluct_222},~\ref{v_fluct_222} and ~\ref{w_fluct_222}, plotted over multiples of the viscous boundary layer thickness $\delta_\nu$ in Figure~\ref{fig:rms_R}. The dashed line indicates the estimated viscous boundary layer thickness according to equation~\eqref{eq:vicous_boundary_layer}. The profiles for the horizontal velocity fluctuations $u^\prime$ (blue) and $w^\prime$ are virtually identical up to approximately three viscous boundary layers from the wall and show the typical profile as reported in reference~\cite{zhang_statistics_2017}.  Furthermore, the position of the profile maxima, which can be used to estimate boundary layer thickness~\cite{scheel_thermal_2012}, is close to the boundary layer thickness derived from scaling theory. The plate-normal or vertical fluctuation profile $v^\prime$ (black) differs from the horizontal fluctuation profiles due to the absence of a main shear flow necessary for a Blasius-type boundary layer to emerge. The Blasius-type boundary layer has been found to be model RBC convection with a dominant LSC~\cite{sun2008experimental}.

\subsection{Inferred temperature and convective heat transfer}
\begin{figure}[H]
\centering
\includegraphics[width=0.7\textwidth]{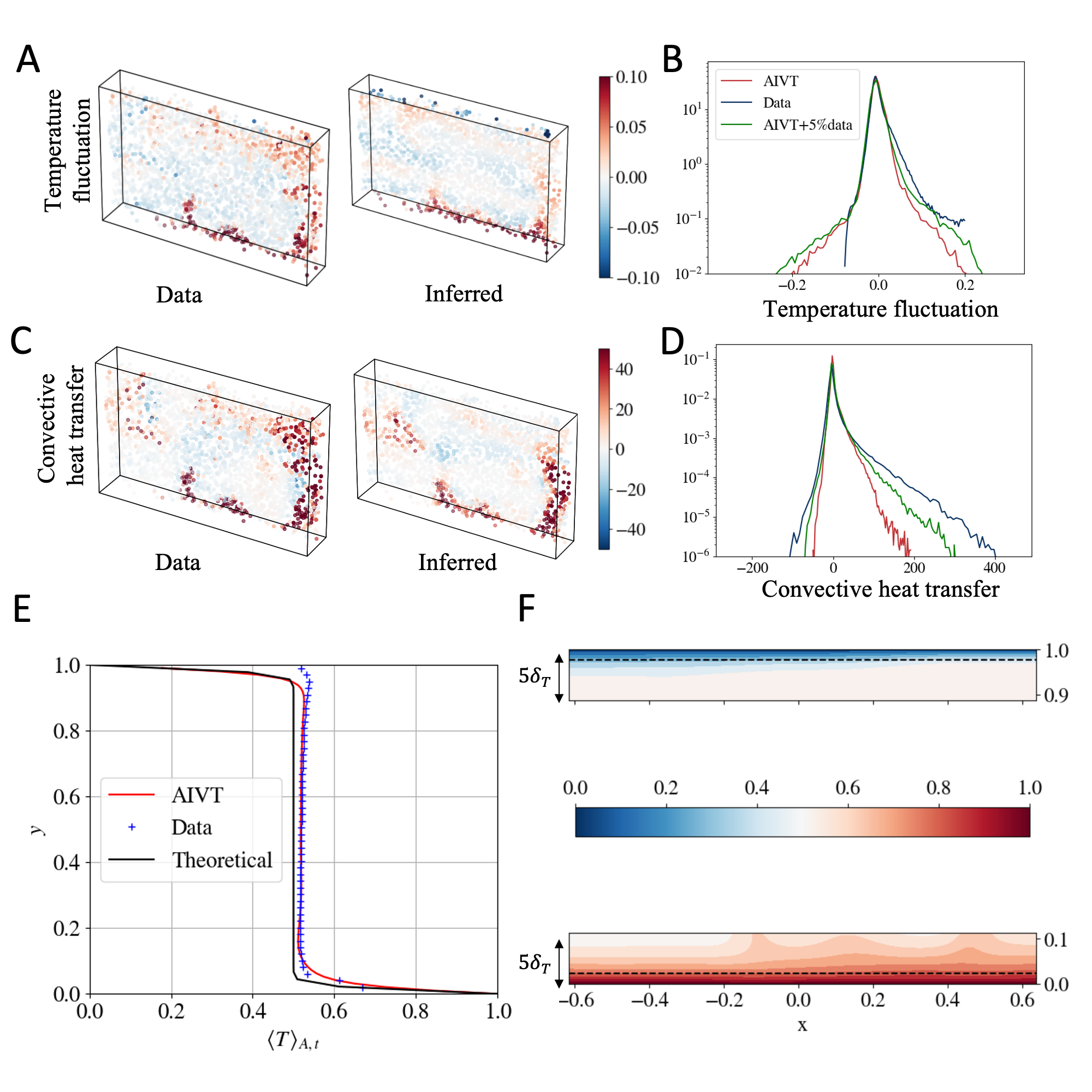}
\caption{\textbf{Inferred temperature and heat transfer results.} (\textbf{A}) Comparison of the scatter plot of the measured and inferred temperature fluctuations at particle positions for an exemplary snapshot. (\textbf{B}) PDFs of the measured (blue), inferred (red), and reconstructed (green) temperature fluctuations at the particle locations. Due to the nonuniform sensitivity of the TLCs, the measurable temperature range is not symmetric but shifted towards positive values. (\textbf{C}) Comparison of the scatter plot of the measured and inferred convective heat transfer at particle positions for an exemplary snapshot. (\textbf{D}) PDFs of the measured (blue), inferred (red), and reconstructed (green) convective heat transfer at the particle locations. PDFs are skewed towards positive values since overall heat is transferred from the bottom to the top. The tails of the measured convective heat transfer PDF are extended further. We attribute this to the measurement uncertainties, inevitably leading to wider PDFs and the tendency of the PIKANs to produce smoothed results. Including a few temperature observations in the training process shifts the results closer to the measurements. (\textbf{E}) Vertical profile of mean temperature $\langle T \rangle _{A,t}$. The red line denotes the inferred profile, the blue markers the profile obtained from the binned Lagrangian measurement data, and the black line the temperature profile as proposed in reference~\cite{shishkina_mean_2009}. Since measured and inferred profile collapse in the bulk region, the offset of the bulk temperature from the theoretical profile is likely caused by slight deviations from the idealized Boussinesq approximation case, as shown in reference~\cite{horn_rotating_2014}.
Close to the heating and especially the cooling plate, temperature measurements become unreliable due to the limited temperature range of the TLC particles. (\textbf{F}) View of the temperature field in the top and bottom boundary region covering the distance of 5 $\delta_T$ from the respective plate. The thermal boundary layer thickness, $\delta_T$, is indicated as a black dashed line. The color bar for both plots is presented in the middle of the figure.}
\label{fig:temperature_results}
\end{figure}

As outlined in the problem description, AIVT's main advantage is that it estimates hard-to-measure quantities like temperature from more easily experimentally accessible quantities like velocity. To validate our model's temperature predictions, we use all the available experimental data and obtain an average relative $L^2$ error of 3.62\% with an uncertainty of less than 1\% (See Table~\ref{Results_apendix}). In Figure~\ref{fig:temperature_results}(A), we compare exemplary snapshots of the measured and inferred temperature fluctuation at the particle positions. The direct comparison shows that the AIVT approach correctly infers the detaching thermal plumes and temperature structures. However, the thermal plumes appear more pronounced in the scatter plot of the measures of temperature fluctuations.

To analyze the inference quality statistically, we computed the PDFs of the measured (blue) and inferred (red) temperature fluctuations, which are shown in Figure~\ref{fig:temperature_results} (B). This plot additionally features the PDF of the temperature fluctuations reconstructed by the AIVT model when trained on $5\%$ of the velocity and temperature data (green). Most noticeably, the PDF of the measured temperature is not symmetric around 0 as one would expect in RBC within the Boussinesq approximation regime but shifted towards higher values. This is related to the nonuniform temperature sensitivity and limited range of the TLCs. Comparing the PDFs of the measured and inferred temperature, we see that the match for values is close to 0 but deviates towards the tails. On the one hand, this might be related to the bias of the measurement technique towards higher values and measurement uncertainties that spread the PDFs. On the other hand, this could also be caused by a slight smoothing of the AIVT model. This is further supported by considering the PDF of the reconstructed temperature, whose positive tail is more closely aligned with the measured temperature, suggesting that by feeding only a few temperature data points, the smoothing of the AIVT model can be reduced.

Since the joint measurement of temperature and velocity enables us to calculate the convective heat transfer directly, we present snapshots of the measured and inferred convective heat transfer at the particle positions in Figure~\ref{fig:temperature_results}(C). The scatter plots show the same time instance as the scatter plot of the temperature. Looking at the scatter plots, we observe the increased convective heat transfer associated with the thermal plumes. When comparing the two scatter plots, we see the strong similarity in the structures even though the high convective heat transfer region at the top right of the measured results is not as pronounced in the inferred results. In both cases, regions of negative local heat transfer are visible. While in RBC, generally positive temperature fluctuations and positive vertical velocity as well as cold temperature fluctuations and negative vertical velocity are correlated, there are regions where, due to viscous forces, adjacent colder fluid is dragged upwards or hot fluid downwards, respectively~\cite{moller2022combined}. This shows that the AIVT model is capable of inferring these regions. 

Similar to the temperature, we present the PDFs of the convective heat transfer in Figure~\ref{fig:temperature_results}(D) to obtain an overall measure of the inference quality. The plot shows the PDFs of the measured (red), inferred (blue), and reconstructed (green) heat transfer. We observe that all PDFs are skewed towards positive values since overall heat is transferred from the bottom to the top of the system. The PDF of the inferred convective heat transfer shows the highest probability for small-magnitude convective heat transfer events. Besides that, the curves match closely in the vicinity of $J=0$ but differ regarding the positive and negative tails. Here, we see that the PDF of the measured convective heat transfer shows the widest spread, followed by the PDF of the reconstructed convective heat transfer, while the inferred heat transfer shows the least spread. This is consistent with our observations from the PDFs of the temperature fluctuations and the scatter plot of the convective heat transfer. Again, we observe that the inferred AIVT results are smoothed compared to the measurement data and that by including a few temperature observations, the reconstructed convective heat transfer matches the experimental observation more closely.

Another important aspect is the vertical profile of the temperature $\langle T\,\rangle _{A,t}$ averaged over time and along the horizontal directions. Therefore, in Figure~\ref{fig:temperature_results}(E), we compare the inferred vertical temperature profile (red), the temperature profile from the binned measurement data (blue crosses), and the theoretical temperature profile (black) as proposed by Shishkina and Thess~\cite{shishkina_mean_2009}. By contrasting the profiles, we observe a close match between the binned measurements and the inferred temperature profile in the bulk region. In comparison to the theoretical temperature profile, the measured and inferred temperature profiles show slightly higher temperatures in the bulk region. Since this trend is prevalent in the measured and inferred temperature, we attribute this to a minor deviation of the experimental configuration from the idealized Boussinesq approximation model, which has been shown to result in an increase of the bulk temperature~\cite{horn_rotating_2014}. Close to the boundary, the temperature measurements become unreliable, especially at the cooling plate, hence the steep temperature gradient is not captured by the measurements. Close to the heating plate, where the temperature range is sufficient to capture parts of the temperature gradients, we see that the inferred temperature and measured temperature profile match. The AIVT model is able to infer the temperature close to the wall, and the inferred and theoretical temperature profiles collapse in the cold boundary region. For the hot boundary region, the inferred temperature profile shows a less steep gradient but follows the data. Hence, the deviation from the theoretical profile may be caused by the constrained region of interest and time span, as well as the presence of detaching thermal plumes, which have been shown to extend the thermal boundary layer locally~\cite{tegze_three-dimensional_2024}.

One of the main advantages of physics-informed machine learning models is that the predictions can be evaluated at any arbitrary point inside the domain. Thus, we use the AIVT model to infer the temperature in the boundary regions that cannot be visualized from the experimental observations. Figure~\ref{fig:temperature_results}(F) shows the temperature in the top and bottom boundary region up to 5 thermal boundary layer thicknesses $\delta_{T}$ in the $x-y$ plane at $z=0$. The thermal boundary layer thickness is estimated according to equation~\eqref{eq:T_boundary_layer} with the Nusselt number $Nu=22$ being estimated based on the Grossmann-Lohse theory~\cite{stevens_unifying_2013} and is indicated by the horizontal black dashed line. In the hot boundary region, we observe the presence of steep temperature gradients beyond the indicated theoretical boundary layer thickness and the footprint of the detaching hot thermal plumes, which are visible in the scatter plots of the temperature and convective heat transfer. This is consistent with the results by Tegze and Podmaniczky~\cite{tegze_three-dimensional_2024}, who showed that detaching thermal plumes locally extends the thermal boundary layer further into the bulk.
For the cold boundary region, we see a close match of the steep temperature gradient with the boundary layer thickness. 

\begin{figure}[H]
\centering
\includegraphics[width=\textwidth]{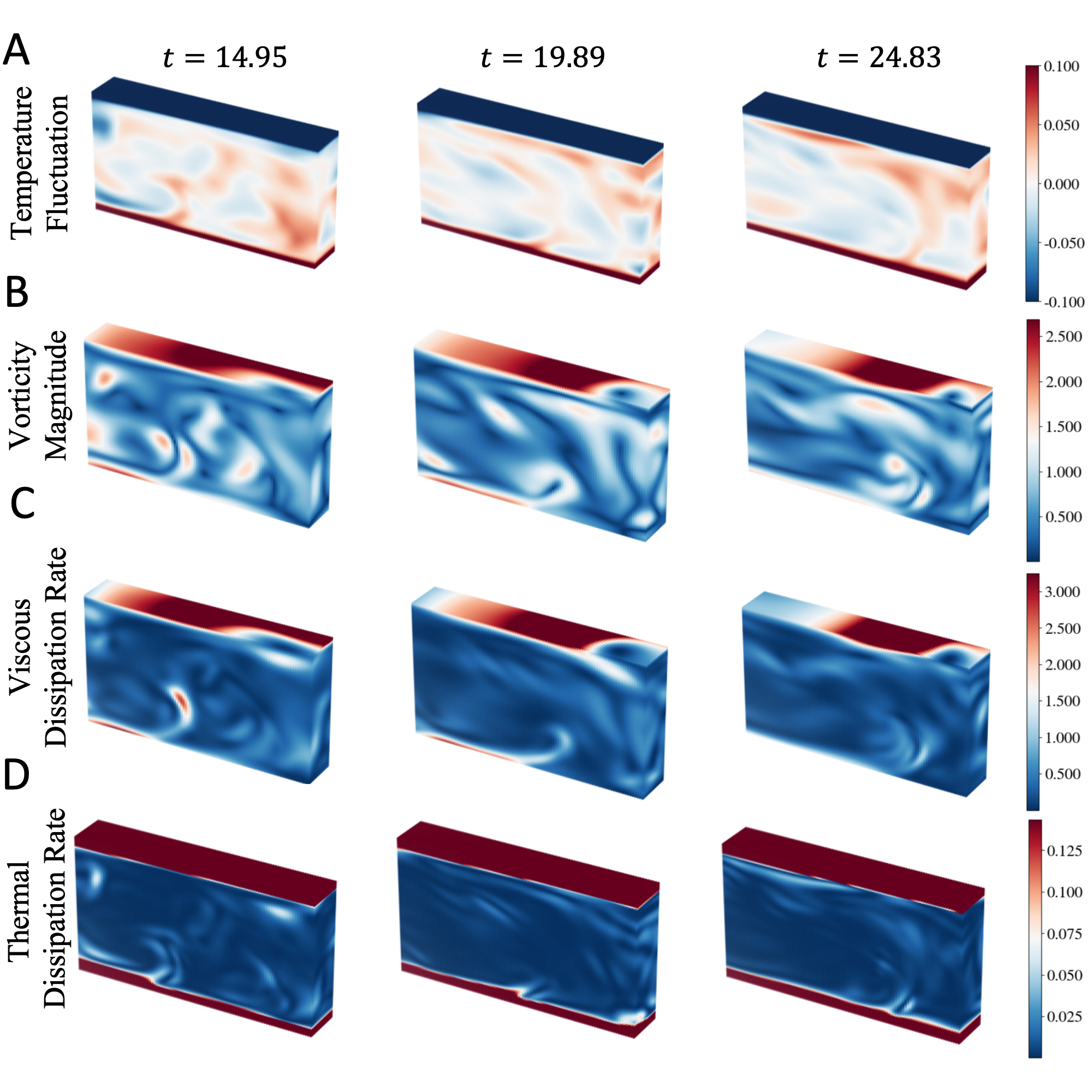}
\caption{\textbf{Inferred turbulent fields}. Three snapshots of the inferred temperature fluctuations (\textbf{A}), vorticity magnitude (\textbf{B}), viscous dissipation rate (\textbf{C}), and thermal dissipation rate (\textbf{D}). The snapshots show the development of the flow over 15 free-fall time units. Snapshots like this are usually only available from DNS. Note the correlation between features among the different quantities.}
\label{fig:snapshots}
\end{figure}
To further analyze the thermal boundary layer, we show the normalized root-mean-squared  
($\sigma/(\sigma)_{\max}$) temperature fluctuation profile plotted over multiples of the thermal boundary layer thickness in Figure~\ref{fig:rms_R} (B). The profile shows the distinct shape also reported by others~\cite{lui_spatial_1998,zhou_thermal_2013}.
The position of the profiles maximum, which is commonly used to estimate the boundary layer thickness~\cite{scheel_thermal_2012}, almost perfectly matches the thermal boundary layer thickness estimated from scaling laws~\cite{grossmann_scaling_2000,stevens_unifying_2013}, showing that inferred temperature results are consistent with well-established and proven theory. These results demonstrate that the AIVT model is capable of inferring the temperature in the vicinity of the thermal boundary layers that are physically sound, consistent with theory and results published by others, purely from Lagrangian velocity data.

\subsection{Time series of instantaneous fields}
Due to the continuous and differential representation of the flow by AIVT, we are not limited to reconstructing the velocity and inferring the temperature but furthermore are able to infer quantities requiring the gradients of these fields like vorticity, the viscous dissipation rate, and the thermal dissipation rate. To this end, we present in Figure~\ref{fig:snapshots} high-resolution snapshots of the inferred temperature (A), vorticity magnitude (B), viscous dissipation rate  $\epsilon_K$ (C), and thermal dissipation rate  $\epsilon_T$ (D) for three different time instances. Note that simultaneous high-resolution snapshots of all the quantities were previously exclusive to direct numerical simulations. Comparing the fields of various quantities, we observe the imprint of the thermal plumes, which are associated with high kinetic energy and thermal variance. We observe high values of the vorticity magnitude and viscous dissipation rate close to the top plate. This is likely caused by the impact of the LSC, see Figure~\ref{fig:velocity_results}(B), resulting in strong shear and, thereby, high values of vorticity and viscous dissipation. The highest thermal dissipation occurs in the thermal boundary layers. Furthermore, the edges show that the thermal plumes are visible due to an increased thermal dissipation rate. In contrast to viscous dissipation, regions of high thermal dissipation appear more confined, which is typical for Pr~$>$~1.

\subsection{Statistics of gradient-based quantities}
\begin{figure}[h]
\centering
\includegraphics[width=0.7\textwidth]{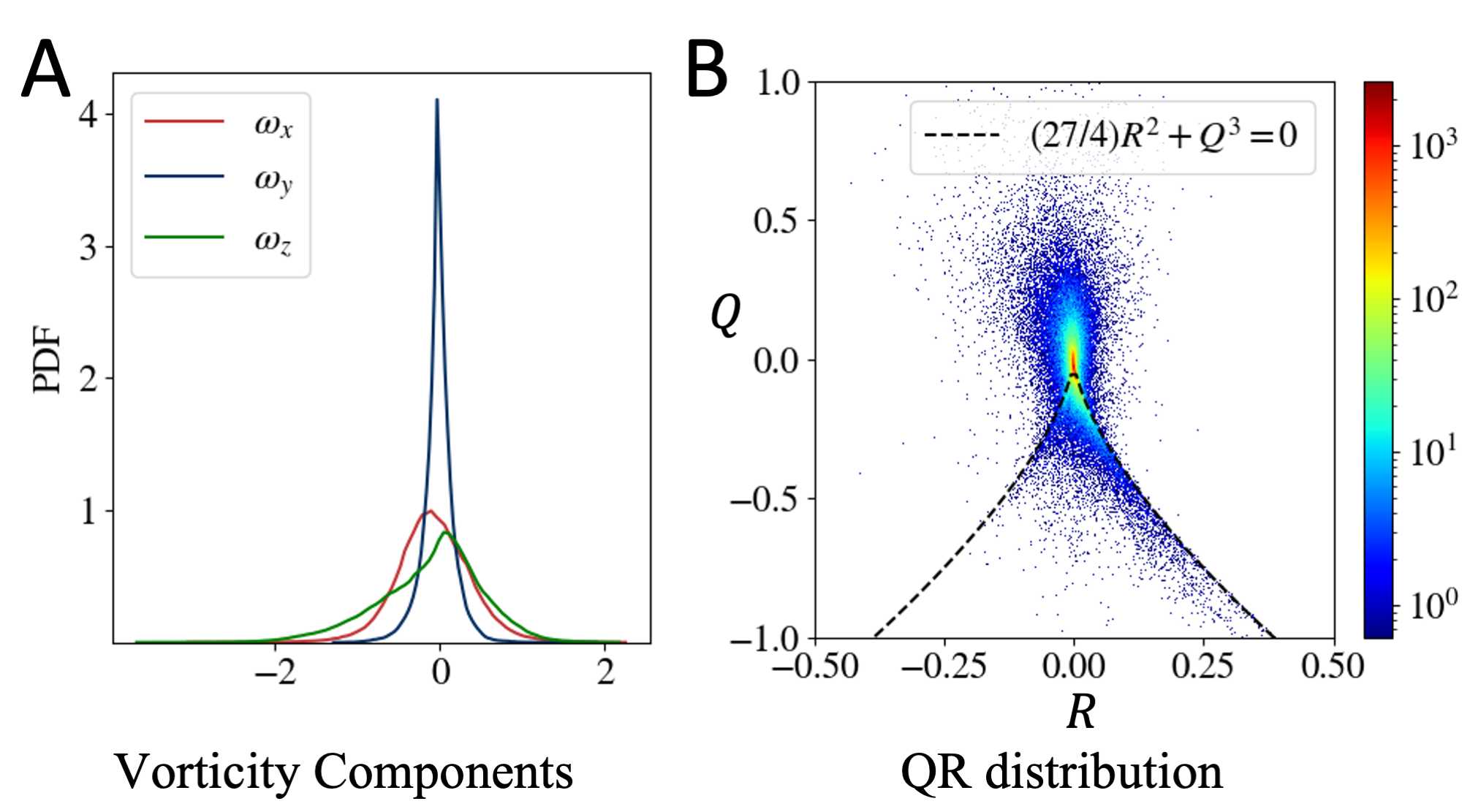}
\caption{\textbf{Velocity gradient-based statistics.} (\textbf{A}) PDFs of the $\omega_x$ (red),$\omega_y$ (blue), and $\omega_z$ (green) components. $\omega_x$ and $\omega_y$ show Gaussian-like distributions, with the distribution of $\omega_y$ having a smaller standard deviation; $\omega_z$ shows a similar width as $\omega_x$ but is slightly skewed. (\textbf{B}) Joint distribution of the $Q$ and $R$ invariants of the velocity gradient tensor. The green line indicates the Vieillefosse tail $(27/4)R^2 +Q^3 =0$. The joint PDF shows the typical elongated and sheared drop shape with the highest probability of low $Q$ and $R$ events and preference towards the right branch of the Vieillefosse tail. This shape is common for many turbulent flows as reported by~\cite{meneveau_lagrangian_2011,johnson_multiscale_2024}.}
\label{fig:vel_grad_pdf}
\end{figure}

Since we have shown that the AIVT model provides meaningful results for gradient-related quantities of temperature and velocity, we now validate the AIVT model further by studying the statistics of these quantities. Therefore, we plot the PDFs of the $\omega_x$ (red), $\omega_y$ (blue), $\omega_z$ (green) components in Figure~\ref{fig:vel_grad_pdf}(A). The PDF of $\omega_y$ shows a Gaussian shape and is closely distributed around 0, indicating little rotation around the vertical axis. In contrast, the PDFs of $\omega_x$ and $\omega_z$ show a wider distribution, indicating higher rotation magnitudes along the horizontal axis. This coincides with the rotation of the LSC. While the $\omega_x$ PDF shows almost a Gaussian shape, the negative tail of the $\omega_z$ PDF decreases almost linearly.

As another way to check for the consistency of the inferred gradient of the velocity field, we show the joint PDF of the $Q$ and $R$ invariants of the velocity tensor in Figure~\ref{fig:vel_grad_pdf}. The dashed line indicates the Vieillefosse tail $(27/4)R^2 +Q^3 =0$. $Q$ and $R$, which are defined in equation~\ref{eq:invariants}, represent the balance between enstrophy and strain-rate-squared magnitude, and between enstrophy production and dissipation production. Thereby, the joint PDF of $Q$ and $R$ is a concise way to represent information about the velocity gradient~\cite{johnson_multiscale_2024}.
The joint PDF shows the typical elongated drop shape with the highest joint probability at $Q=R=0$ and a preference towards the right Vieillefosse tail.
This typical shape has been observed for many turbulent flows~\cite{meneveau_lagrangian_2011,johnson_multiscale_2024} and proves that inferred velocity gradients are consistent with the theory.

\begin{figure}[H]
\centering
\includegraphics[width=0.9\textwidth]{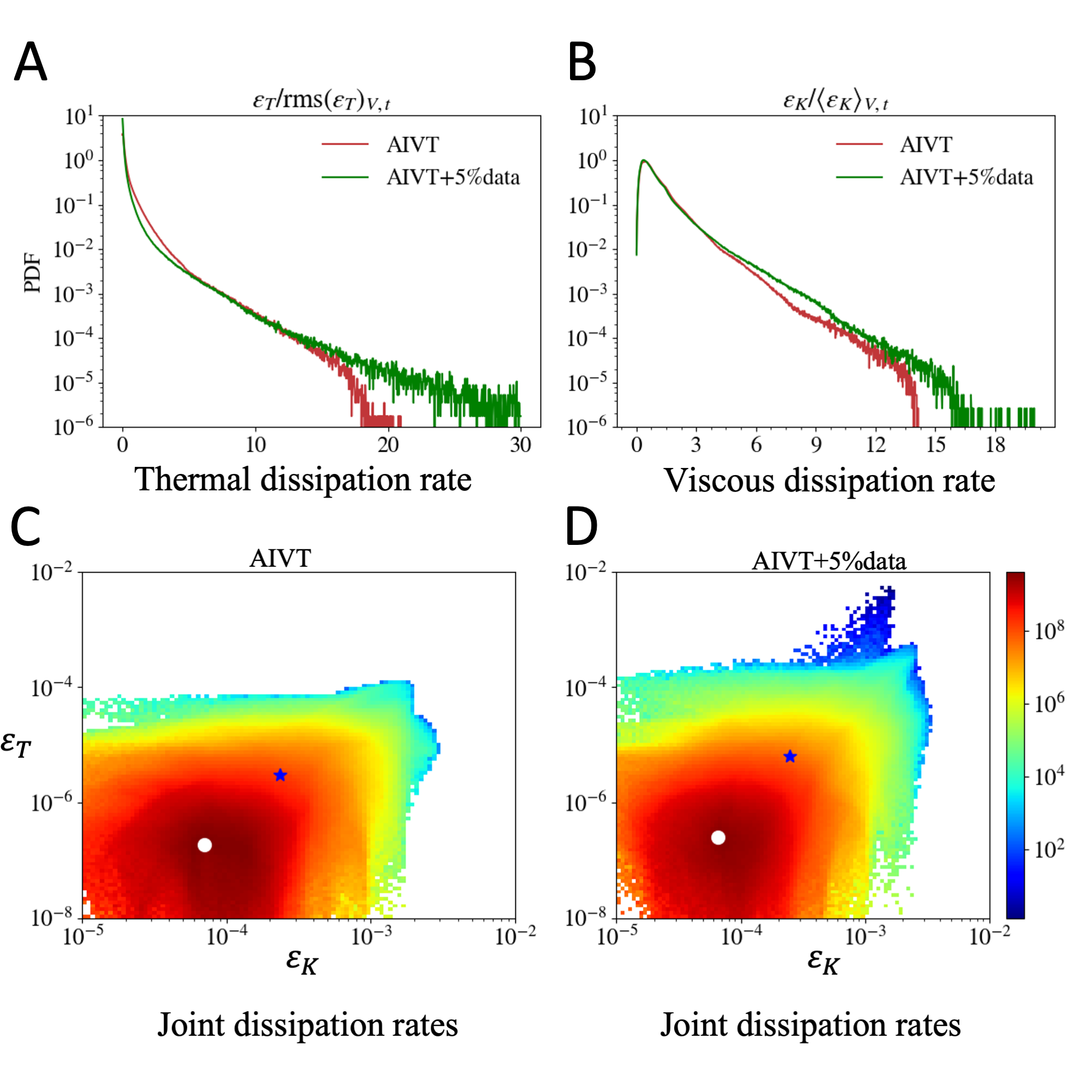}
\caption{\textbf{Temperature and velocity dissipation rate statistics.} PDFs of the normalized thermal (\textbf{A}) and viscous (\textbf{B}) dissipation rates in the center region 0.3~$<y<$~0.7 for the AIVT model trained without (red) and with 5$\%$ temperature data (green). The PDFs of the thermal and viscous dissipation rates are similar to the results obtained from point-wise measurements of the thermal dissipation rate~\cite{he2007measured} and to the results obtained from local measurements of the viscous dissipation rate~\cite{xu_experimental_2024}, respectively. Including sparse temperature observations into the training nudges the model towards higher magnitude events.\newline
Joint PDF of the thermal $\epsilon_T$ and viscous dissipation rates $\epsilon_K$ in the core region 0.3~$<y<$~0.7 of the domain obtained from the AIVT model trained only on velocity (\textbf{C} and additional few temperature observations (\textbf{D}). The white cross and the blue star denote the most probable joint events of $\epsilon_T$ and $\epsilon_K$ and the blue star the intersection of $\langle \epsilon_T \rangle _{V,t}$ and $\langle \epsilon_K \rangle _{V,t}$, respectively. The events with the highest joint probability have lower values of $\epsilon_T$ and $\epsilon_K$ than the respective mean values of $\epsilon_T$ and $\epsilon_K$. This and the overall shape of the distribution are similar to the results obtained from DNS of RBC by Kaczorowski and Xia~\cite{kaczorowski_turbulent_2013}.}
\label{fig:dissipation_plot}
\end{figure}

Since the AIVT model is not constrained by a finite resolution, it provides access to small-scale properties, which previously were notoriously hard to obtain from measurements, especially for LPT with limited particle image densities. Thus, we show the normalized PDF of thermal dissipation rate within the center region 0.3$ <y< $0.7 inferred by the AIVT (red) and by the AIVT model trained on limited temperature data (green) in Figure~\ref{fig:dissipation_plot}(A).
The PDF of the AIVT model purely trained on velocity data shows a higher probability for low thermal dissipation rate events compared to the PDF obtained from the AIVT model trained on limited temperature data whose tail extends further towards high magnitude events.
Again, this confirms that the AIVT model, purely trained on velocity data, tends to smooth but can be guided by including only a few temperature observations. We compare the PDFs with normalized PDFs of point measurements of the thermal dissipation rate at Ra = $8.2 \times 10^8$ and Pr = 5.4 in the center of a RBC cell reported by He et al.~\cite{he2007measured}. Their results show the same trend and closely match the PDF of the AIVT model trained with a few temperature data points. This shows that our results are consistent with results obtained by other research while achieving unprecedented spatial resolution.

We proceed similarly for the viscous dissipation rate and show the normalized PDF of viscous dissipation rate within the center region 0.3$ <y< $0.7 inferred by the AIVT (red) and by AIVT model trained on limited temperature data (green) in Figure~\ref{fig:dissipation_plot}(B). We observe that both PDFs show the same trend and match closely up to a normalized viscous dissipation of $\approx$ 4. Beyond this locaion, the PDF of the AIVT model trained on velocity data only attributes a lower probability to high viscous dissipation events. This aligns with the previously observed underestimation of high-magnitude events. Interestingly, the comparison shows that also velocity gradient-related quantities benefit from including temperature data in the training process.
To validate the results, we compare the viscous dissipation rates results of local measurements at the center of a RBC cell at Ra = $2 \times 10^8$ and Pr = 4.34, which were recently published by Xu et al.~\cite{xu_experimental_2024}.
While the overall trend aligns with the literature results, the reported PDFs extend further toward high dissipation events. The difference might be, to some extent, due to the difference in Ra number and, more importantly, in Pr. Since the Prandtl number is more than two times as high, momentum is more efficiently dispersed, resulting in viscous dissipation events of smaller magnitude.

The proposed method further allows us to investigate the joint variation of the two spatially resolved dissipation rates, a task previously only feasible with DNS. To demonstrate that, we plot the joint PDFs of the thermal and viscous dissipation rate obtained from the AIVT model in Figure~\ref{fig:dissipation_plot}(C) and from the AIVT model trained on limited temperature data~\ref{fig:dissipation_plot}(D). The white dot and the blue star indicate the highest joint probability and the intersection of the individual mean of the thermal and viscous dissipation, respectively. The comparison of the joint PDFs shows a similar shape, albeit the joint PDF obtained from the AIVT model trained only on velocity tends to be offset towards lower thermal dissipation events. Additionally, the AIVT model trained with few temperature observations shows a tail of high thermal dissipation events, albeit with a very low probability. Due to the low probability of these events, they might be caused by outliers in the temperature data that propagate through into the AIVT predictions. The relative positions of the highest probability and the intersection of the individual mean dissipation rates are similar in both joint PDFs. 

Since these results were so far only obtainable from DNS, we compare our results with joint PDFs of the dissipation rates obtained from DNS of RBC at Ra = \{$5\times10^6$, $1\times10^9$\} at Pr = 4.38 reported by Kaczorowski and Xia~\cite{kaczorowski_turbulent_2013}. Contrasting the reported joint PDF with our results, we can see that the general shape of the PDFs matches; however, they are more closely aligned to the PDF reported for Ra~=~$1\times10^9$. While the position of the intersection of the individual mean of the dissipation rate matches, the highest joint probability is shifted by an order of magnitude towards a higher thermal dissipation rate in our results. This and the presence of events at low $\epsilon _K$ and thermal dissipation rate of $\epsilon _T \approx 10^{-4}$, especially in the joint PDF obtained from the model trained on temperature data, might be related to the difference in Pr resulting in higher thermal dissipation events.
Once more, this shows that our AIVT model enables us to infer high-fidelity DNS-level fields purely from velocity measurements, which can be further improved by incorporating sparse temperature observations. 

\newpage
\section{Discussion and concluding remarks}
\label{Summary}
  In this manuscript, we propose a novel AIVT method capable of inferring DNS-level solutions and statistics of temperature, velocity, and their gradients at arbitrary resolutions purely from Lagrangian velocity measurements. To validate our model, we computed the relative $ L^2$ errors against experimental observations and compared the AIVT-predicted flow statistics with the measurement data and previous studies. 

The proposed AIVT method is comprised of four components that enhance the baseline PINN model, namely using a cKAN as a representation model instead of MLP, reformulating the PDE into the velocity-vorticity formulation(VV), and two optimization algorithm improvements, namely, RBA-R, and sequential training. While PINNs have shown remarkable performance on simulated data of laminar flows~\cite{raissi2020hidden}, the potential of the AIVT model lies in its capability to infer hidden fields from experimental turbulent data. In the Table~\ref{ablation}, we present an ablation study demonstrating that our proposed modifications and extensions to previous approaches~\cite{cai2021artificial, boster2023artificial} are crucial for obtaining the results presented in this manuscript.

We trained the model using a unique set of joint Lagrangian temperature and velocity measurements, which allows us to compare the reconstructed velocity, inferred temperature, and convective heat transfer with the validated experimental data~\cite{kaufer_volumetric_2024}. 
Our results show that the reconstructed velocities are closely aligned (i.e., approximately $10\%$ error) with the measured data, demonstrating the capabilities of the AIVT model to assimilate Lagrangian velocity information. Additionally, AIVT can successfully infer the temperature (i.e., approximately $4\%$ error), convective heat transfer, and gradients of the temperature and velocity field purely from experimental velocity data, governing equations, and boundary conditions. 

The direct comparison between the inferred and measured temperature field and statistics unveiled that AIVT captures the flow features, and the increased heat transfer is associated with the thermal plumes and local region of negative convective heat transfer. The inferred convective heat transfer PDF shows the typical skew toward positive values, albeit with a tendency to underestimate the event's magnitude. Nevertheless, AIVT provides a physically consistent result of temperature and velocity, which can be further fine-tuned by including only a few temperature observations in the process.

The analysis of the vertical temperature profile showed a collapse of the measured and inferred results in the bulk region and good agreement with the theoretical results in the regions close to the plates in which the temperature measurements are unreliable. Our study of inferred snapshots of temperature in the hot and cold boundary regions showed an agreement on the theoretical boundary layer thickness, especially in the cold boundary region, which is further verified by the vertical profile of the temperature fluctuations that matches literature results~\cite{lui_spatial_1998,zhou_thermal_2013}. In the hot boundary region, the imprint of detaching thermal plumes was successfully reconstructed~\cite{tegze_three-dimensional_2024}. We demonstrated the AIVT capabilities to provide results of DNS-level fidelity of various quantities by presenting snapshots of the temperature fluctuation, vorticity magnitude, and viscous and thermal dissipation rates.

Our study of the PDFs of the vorticity components unveiled enhanced vorticity magnitude along the horizontal axes, which is consistent with the orientation of the LSC. The joint distributions of the $Q$-$R$ invariants of the velocity gradients tensors showed the well-known elongated drop shape common to many turbulent flows~\cite{meneveau_lagrangian_2011,johnson_multiscale_2024}.

Our comparison of the PDFs of the thermal and viscous dissipation rates with experimental results reported by He et al.~\cite{he2007measured} and Xu et al.~\cite{xu_experimental_2024}, respectively, showed qualitative agreement.
Note here that their results were obtained in separate experiments at slightly different parameters by highly specialized pointwise or localized measurement. In contrast, our method provide results over the full domain height of the various quantities at arbitrary spatial-temporal resolution. 
Such results were previously exclusive due to DNS. 
This is further underlined by the joint PDFs of the thermal and viscous dissipation rate, which we compared to DNS results reported by Kaczorowski and Xia~\cite{kaczorowski_turbulent_2013}. Our results show an overall similar distribution, and differences might, to some extent, be explained by slight differences in Rayleigh and Prandtl number. As for the convective heat transfer, the results can be further improved by including a few temperature observations in the training process. These can be easily achieved by standard thermocouples or resistance thermometers.

Our future research aims to improve the AIVT model further and better infer high-magnitude events. Secondly, we aim for a deeper integration of measurement data processing and scientific machine learning instead of treating them as different steps, thereby potentially benefiting from symbiosis effects. Finally, we want to adapt our method to various other fields, measurement methods, and experimental data, e.g., magneto-hydrodynamics or solid mechanics.

In summary, we proposed and applied a method that represents a paradigm shift in fluid mechanics research by fusing cutting-edge flow measurement techniques with the potential of the latest scientific machine-learning methods. In combination, they can potentially change our perspective on experimental and computational methods.
We think that the concept of fusing experiments and scientific machine learning is not limited to thermal convection but can be extended to fluid mechanics in general and even beyond, where the benefit of combining experimental data and scientific machine learning opens up new research perspectives.

\section*{Acknowledgements}

J.D.T., Z.W., and G.E.K. acknowledge support from NIH grant R01AT012312, the MURI-AFOSR FA9550-20-1-0358 project, and the ONR Vannevar Bush Faculty Fellowship (N00014-22-1-2795). Additionally, G.E.K. and M.M. are supported by the DOE SEA-CROGS project (DE-SC0023191).

The work of T.K. and C.C. was supported by the Carl Zeiss Foundation within project no. P2018-02-001 “Deep Turb – Deep Learning in and of Turbulence," by the DFG Priority Program SPP 1881 on “Turbulent Superstructures” within project no. 429328691 and project no. 467227170.
\newpage
\appendix 
\section{Detailed Results}
\label{Result_Det}
\begin{table}[h]
\footnotesize
\centering
\caption{Results}

\begin{tabular}{lcccc}
Method & $RL^2$ $u$(\%) &$RL^2$ $v$ (\%)&  $RL^2$ $w$(\%)& $RL^2$ $T$ (\%)\\
\midrule
AIVT & $9.68\pm0.23$ & $10.86\pm0.073$& $12.00\pm0.13$&$3.62\pm0.89$\\
%AIVT+5\%Data & 9.75 & 10.93& 12.13&2.77 \\
AIVT+5\%Data & 9.94 & 10.01& 12.32&2.76 \\
\bottomrule
\end{tabular}
\caption{(First row) Average Relative $L^2$ error for the proposed method (AIVT) along with the uncertainty (i.e, three standard deviations) from 5 independent runs. (Second row) Results of an AIVT model trained with 5\% velocity and temperature data.}
\label{Results_apendix}
\end{table}

\subsection{Reconstructed Velocity}
\begin{figure}
\centering
\includegraphics[width=\textwidth]{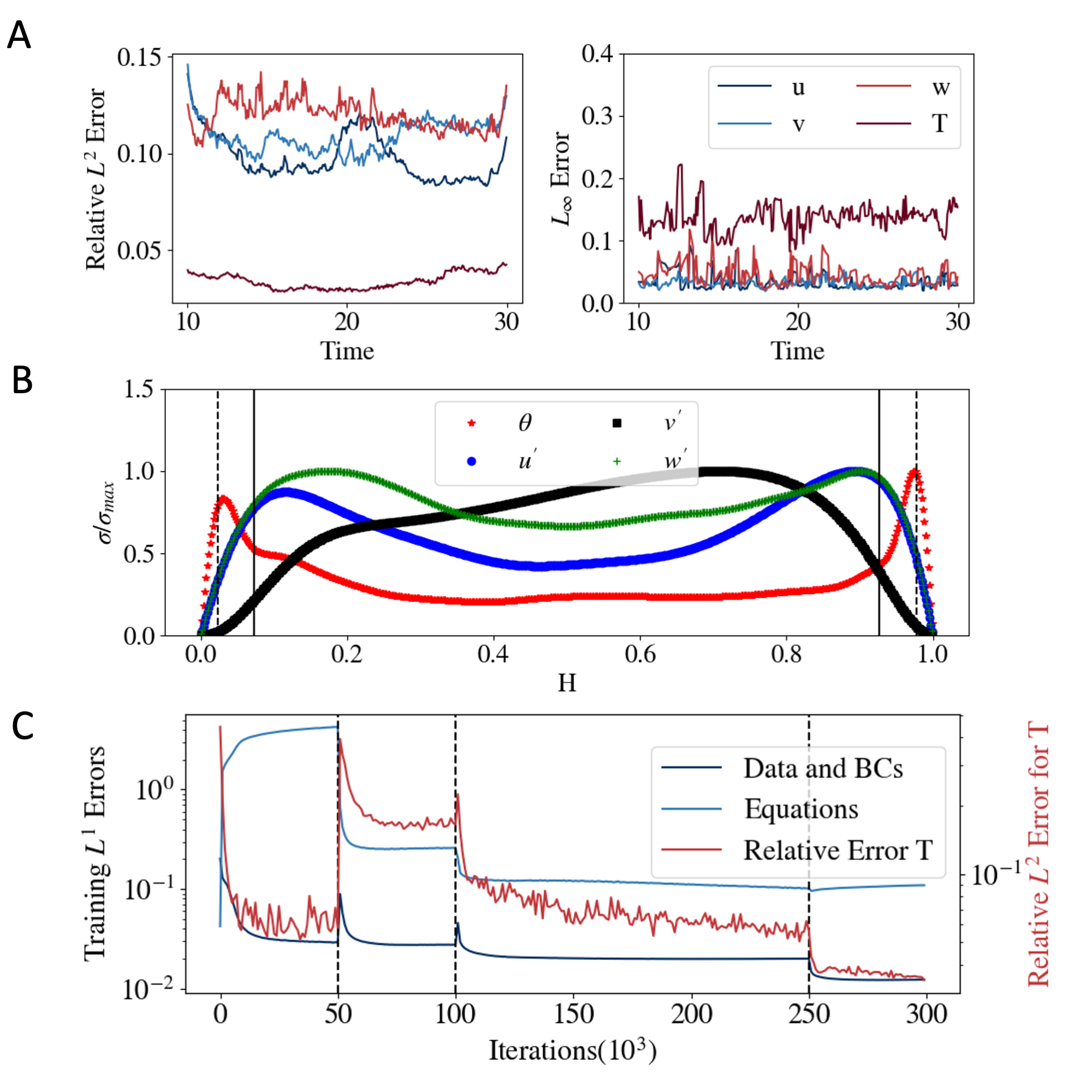}
\caption{\textbf{Compiled Results:} (A) Relative $L^2$ and $L^{\infty}$ (maximum absolute difference) for the analyzed timesteps. (B) Normalized root-mean-squared  
($\sigma/(\sigma)_{\max}$) velocity $u^\prime, v^\prime, w^\prime$ temperature $T^\prime$ fluctuations on the whole domain. The vertical dashed lines indicate the viscous $\delta_{\nu}$ and thermal $\delta_T$ boundary layer thicknesses, respectively. (C) Sequential training approach. Initially, we reduced the problem to a data-driven approach, where we learned velocity information and boundary conditions. In the second step, we learn our equations with a lower Ra number. In the third stage, we learn about our full model. In the last state, we refine the solutions in the final stage by changing our loss exponent $q$.}
\label{all_results}
\end{figure}

\begin{figure}
\centering
\includegraphics[width=0.9\textwidth]{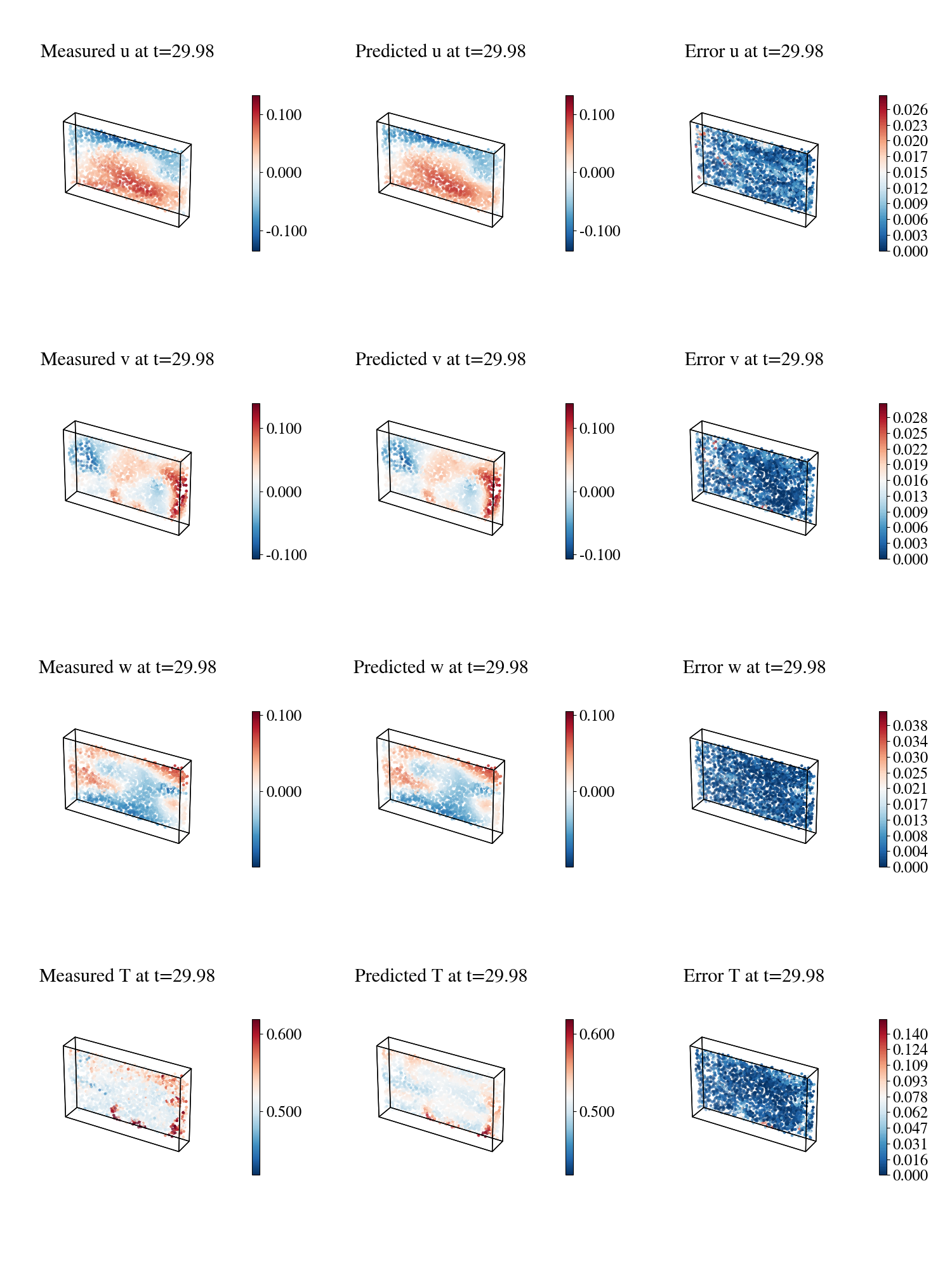}
\caption{Measured and predicted flow fields and their corresponding absolute error for a representative time step (worst case). (First row) Velocity in $x$ direction. (Second row) Velocity in $y$ direction. (Third row) Velocity in $z$ direction. (Fourth row) Temperature.}
\label{point_wise_error}
\end{figure}

We validate or model performance on the core region ($0.1<y<0.9$) of the validation data (i.e., 50\% of the velocity measurements). The AIVT relative L2 errors on the
core region ($0.1 < y < 0.9$) for five different seeds are presented in Table~\ref{Results_apendix}. Notice that the uncertainty (i.e., three standard deviations) from the model parameters is less than $0.25\%$. Figure~\ref{all_results}(A) shows the best-performing model relative to $L^2$ and $L^{\infty}$ (maximum absolute difference), indicating that the error value is consistent for the analyzed time steps (i.e., 282 frames). The pointwise error distribution for the worst case is shown in figure ~\ref{point_wise_error}.

\subsection{Inferred Temperature} Using AIVT, we reconstruct the temperature field purely from the sparse velocity information. To validate these results, we use all the temperature information collected using the proposed experimental method. The AIVT mean relative L2 error is $3.62\%$, and as shown in Table~\ref{Results_apendix}, the uncertainty due to the model parameters is less than $1\%$.

To improve our framework performance, we introduce a second network that uses 5\% of the velocity and temperature for training and validated with the remaining unsee data. As shown in table~\ref{Results_apendix}, even with 5\% data, we can reduce the temperature relative error to 2.76\%. Also notice that reducing the velocity training data has a small influence on the velocity reconstructions.

To further investigate our results, we compute the normalized root mean squared ($\sigma/\sigma_{max}$) velocity and temperature fluctuations (See Figure~\ref{all_results}(B)) along the whole domain. The vertical dashed black lines indicate the thermal boundary layer thickness, while the solid vertical black lines indicate the viscous boundary layer thickness. 

\begin{figure}
\centering
\includegraphics[width=\textwidth]{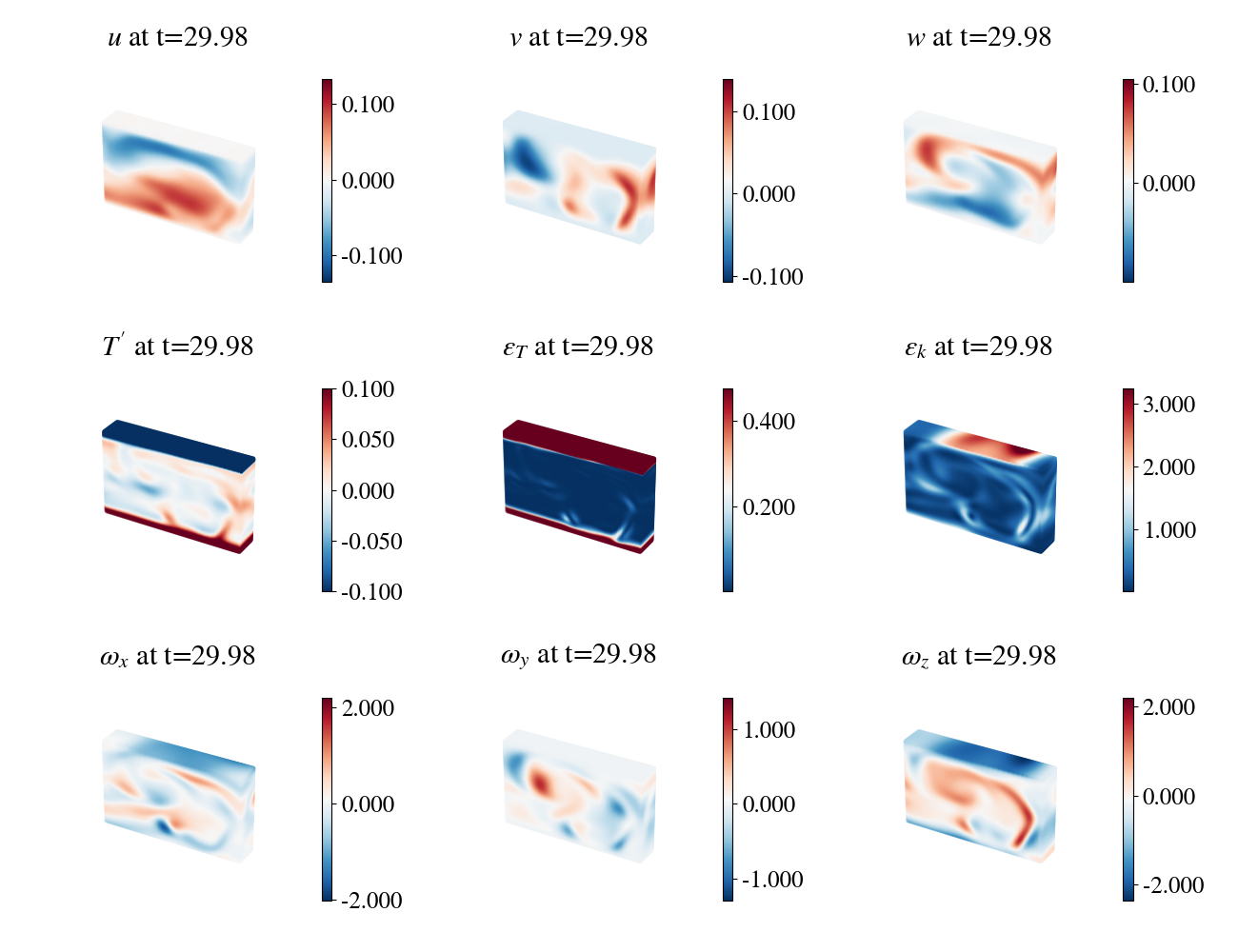}
\caption{Reconstructed 3D velocities and inferred flow fields at a representative time step. (First row) Velocity components in the $x$, $y$ and $z$ directions. (Second row )temperature fluctuations, thermal and viscous dissipation. (Third row) Vorticity components in the $x$, $y$ and $z$ directions.}
\label{reconstruct}
\end{figure}

AIVT outputs continuous and differentiable functions that enable us to define the velocity, temperature, or derived fields (i.e., convective heat transfer, thermal and viscous dissipation rates ) at any point inside the domain. The predictions of these quantities in a dense Eulerian grid are shown in Figure ~\ref{reconstruct} for a representative time.

\subsection{Ablation Study}

The proposed AIVT method is comprised of four components that enhance the baseline AIV model, namely, (1) using a cKAN as a representation model instead of MLP, (2) reformulating the PDE into the velocity-vorticity formulation (VV), and the optimization algorithm improvements: (3) RBA-R and (4) sequential training (Seq). Based on this description, our method can be represented as a composition of each subcomponent, i.e., AIVT=cKAN+VV+RBA-R+Seq. On the other hand, the baseline AIV models ~\cite{boster2023artificial,cai2021artificial} use MLP and the velocity-pressure (VP) formulation (i.e., MLP+VP). The results are shown in Table~\ref{ablation} (A). Notice that using the baseline formulation outperforms the proposed method in the velocity reconstruction; however, it fails to infer the temperature completely. To explore the behavior, we ran an additional MLP model that uses the VV formulation, and we noticed that, even though the velocity reconstruction is affected, the error on the temperature is significantly reduced in Table~\ref{ablation} (B), suggesting that the VV independence of pressure is crucial to infer the hidden temperature.
\begin{table}[ht]
\footnotesize
\begin{tabular}{clcccc}
&Method & $RL^2$ $u$(\%) & $RL^2$ $v$ (\%)&  $RL^2$ $w$(\%)& $RL^2$ $T$ (\%)\\
\midrule
A&MLP+VP (Baseline)\cite{cai2021artificial,boster2023artificial} & 5.74 & 5.70 & 9.82 & 72.48 \\
&cKAN+VV+RBA-R+Seq & 9.66 & 10.86 & 11.95 & 3.42 \\\hline
B&MLP+VV & 10.26 & 14.36 & 12.81 & 41.42 \\\hline
C&MLP+VV+RBA-R+Seq & 14.33 & 17.36 & 15.03 & 6.02 \\
&cKAN+VP+RBA-R+Seq & 5.77 & 4.19 & 9.13 & 5.73 \\
&cKAN+VV+Seq & 9.60 & 12.76 & 11.77 & 4.34 \\
&cKAN+VV+RBA-R & 9.69 & 11.40 & 9.59 & 8.49 \\
\bottomrule
\end{tabular}
\caption{Ablation Study. Relative $L^2$ error comparison between different methods.}
\label{ablation}
\end{table}

To further analyze and interpret the effect of each component of our method, we perform a one-component ablation study in which we remove one component at a time and evaluate the model performance of the ablated system. As shown in Table~\ref{ablation} (C), the most important element in our approach is sequential training, which simplifies the optimization problem by dividing the optimization process into substeps. Also, it can be noticed that RBA-R is crucial to refine the details and achieve a better performance. Finally, notice that the velocity-pressure formulation ($VP$) with sequential training and RBAR outperforms the velocity reconstruction; it cannot be used to learn the temperature. 

To allow a fair comparison between cKAN and MLP-based formulations, we choose an architecture that approximately matches the total number of parameters $|\theta|$, which can be quantified as follows \cite{shukla2024comprehensive}:

\begin{align}
  |\theta|_{MLP}&=L[I+(n_l-1)L+O]\sim \mathcal{O}(n_lL^2)\\
  |\theta|_{cKAN}&=L[I+(n_l-1)Lk+O]\sim \mathcal{O}(n_lL^2k)
\end{align}

\noindent here $I$ and $O$ are the numbers of inputs and outputs, $n_l$ is the number of hidden layer and $L$ is the number of neurons per hidden layer and $k$ is the polynomial order. For this example, cKAN outperforms MLP. However, it is worth noticing that MLP may benefit from wider networks and higher learning rates. Additionally, several MLP enhancements, such as weight normalization or feature expansions, improve the MLP base performance, which cannot be used in the current state of cKAN. 

\newpage

\subsection{Underlying Physical Laws}
\label{Uderlying_Laws}
We consider the flow in the Rayleigh-Bénard convection cell under the Boussinesq approximation (i.e., the full set of the Navier-Stokes Equation). In this study, we use the velocity-vorticity (VV) formulation, which, given its independence of pressure, allows us to infer temperature directly from sparse velocity observations and boundary conditions. The equation  for mass, momentum, and energy conservation and the divergence-free constraints are defined as follows:
\begin{align}
\label{Mom_eq_vw}
\frac{\partial\bm{\omega}}{\partial t}+(\bm{u}\cdot\nabla)\bm{\omega}&=(\bm{\omega}\cdot\nabla)\bm{u}+\sqrt{\frac{Pr}{Ra}}\nabla^2\bm{\omega}+(\nabla\times T \mathbf{\hat{e}}_g)\\
\label{cont_u}
\nabla\cdot\bm{u}&=0\\
\label{cont_vw}
\nabla\cdot\bm{\omega}&=0\\
\label{Temp_Eq}
\frac{\partial T}{\partial t}+(\bm{u}\cdot\nabla)T&=\sqrt{\frac{1}{RaPr}}\nabla^2 T
\end{align}
\noindent where $\bm{u}=(u,v,w)$ is velocity field, $\bm{\omega}=(\omega_x,\omega_y,\omega_z)$ is the vorticity and $T$ is temperature. $\mathbf{\hat{e}}_g$ is a unit vector that defines the direction of gravity. This formulation introduces the vorticity in three dimensions, defined from Equation~\ref{Eq_vorticity} and further constrained using the vector identity described in Equation~\ref{Eq_id_w}. 
\begin{align}
\label{Eq_vorticity}
\bm{\omega}&=\nabla\times \bm{u}\\
\label{Eq_id_w}
\nabla^2\bm{u}&=-\nabla\times \bm{\omega}
\end{align}

The analyzed domain $\bm{x}=(x,y,z)\in\Omega=(-0.6,0.6)\times(0,1)\times(-0.1,0.1)$ is a cuboid located at the center of the Rayleigh-Benard hexagonal cell (See Figure~\ref{fig:VW-KAIV})(A). And the analyzed time domain is comprised of 282 frames corresponding to 20 nondimensional times, i.e., $t\in(10,30)$. 

\subsection{Kolmogorov-Arnold networks}
Kolmogorov-Arnold networks (KANs) are a novel type of neural network inspired by the Kolmogorov-Arnold representation theorem \cite{liu2024kan}. This theorem states that any multivariate continuous function $f(\bm{x}) = f(x_1, x_2, \dots)$ on a bounded domain can be represented as a finite composition of continuous functions of a single variable, and the binary operation of addition. Motivated by this theorem, \cite{liu2024kan} proposed approximating $f(\bm{x})$ as follows:

\begin{equation}
\footnotesize
  f(\bm{\zeta}) \approx \sum_{i_{L-1}=1}^{n_{L-1}} \phi_{L-1,i_L,i_{L-1}} \left( \sum_{i_{L-2}=1}^{n_{L-2}} \cdots \left( \sum_{i_2=1}^{n_2} \phi_{2,i_3,i_2} \left( \sum_{i_1=1}^{n_1} \phi_{1,i_2,i_1} \left( \sum_{i_0=1}^{n_0} \phi_{0,i_1,i_0}(\zeta_{i_0}) \right) \right) \right) \cdots \right)
\label{KAN_net}
\end{equation}

The right-hand side of \eqref{KAN_net} represents a KAN ($KAN(\bm{\zeta})$), where, $\bm{\zeta}=(\zeta_1,\zeta_2,...)$ is the multivariate input, $L$ denotes the number of layers, $\{n_j\}_{j=0}^{L}$ are the numbers of nodes (i.e., neurons) in the $j^{th}$ layer, and $\phi_{i,j,k}$ are the univariate activation functions. The specific form of each $\phi(x)$ defines the variations among different KAN architectures. Among these variations \cite{shukla2024comprehensive} introduced cKANs, which use Chebyshev polynomials as univariate functions. For cKANs, $\phi$ is defined as:
\begin{align*}
 \phi(\zeta,\theta)=w_n\sum_n c_nT_n(\zeta)
\end{align*}
\noindent here, $\theta=(w_n,c_n)$ are trainable parameters and $T_n$ is the $n$-order Chebyshev polynomial defined recursively as $T_n(\zeta)=2\zeta T_n(\zeta)+T_{n-1}(\zeta)$ \cite{karniadakis2005spectral}. Notice that $T_n$ is a composition of the lower order polynomials $T_{n}$ and $T_{n-1}$ which, as described in \cite{shukla2024comprehensive}, is significantly faster than the B-Splines KANs introduced in \cite{liu2024kan}.
\subsection{Artificial Intelligence Velocimetry Thermometry (AIVT)}
\label{AIVT_decription}
AIVT is a scientific machine learning model based on cPIKANs\cite{shukla2024comprehensive} and inspired on AIV\cite{cai2021artificial,boster2023artificial} that can infer and reconstruct flow fields from experimental data and the underlying physical laws. We use AIVT to obtain continuous and differentiable flow and temperature fields from sparse velocity measurements. In particular, we approximate the solutions of the Rayleigh-Bénard equations as follows:
\begin{equation}
  \label{KAN_vw_2}
  (\bm{u}, T') = cKAN(t, \textbf{x}, \theta)
\end{equation}
where $\textbf{x} = (x, y, z)$ represents the inputs namely, $x, y, z$ as the spatial nondimensional coordinates, $t$ as time. $\bm{u} = (u, v, w)$ is the velocity field used to derive the vorticity vector $\bm{\omega} = (\omega_x,\omega_y,\omega_z)$ by applying the curl operator (i.e., $\bm{\omega} = \nabla \times \bm{u}$). This reformulation strictly satisfies Equation~\ref{cont_vw}, as the divergence of the curl of any vector field equals zero. The temperature $T$ is obtained from the predicted temperature fluctuation $T'$ as follows:
\begin{align}
\label{exact_temp_Eq}
 T(t, \textbf{x}, \theta) = g(\textbf{x}) + \varphi(\textbf{x}) T'(t, \textbf{x}, \theta)
\end{align}
Here, $g(\textbf{x})$ is a function that satisfies the boundary conditions, and $\varphi(\textbf{x})$ is a distance function that equals zero at the boundaries. In this study, we define $g(\textbf{x})$ and $\varphi(\textbf{x})$ as follows:
\begin{align}
g(\textbf{x})=g(y)&=T_h-y\frac{(T_h-T_c)}{H}\\
 \varphi(\textbf{x})=\varphi(y)&=(y-H)(y)
\end{align}
\noindent where $T_h=1$ and $T_c=0$ are the hot and cold plate's temperature, and $H=1$ is the height of the Rayleigh-Benard convection cell. As described in \cite{sukumar2022exact}, this formulation allows us to exactly enforce the temperature boundary conditions, with $T|_{y=1} = 1$ and $T|_{y=0} = 0$.

We impose the remaining constraints by optimizing a combined loss function that minimizes the error from data, boundary conditions, and equations. The data loss ($\mathcal{L}_D$) controls the mismatch between the network prediction and experimental observations and is explicitly defined as:
\begin{align}
\label{data_eq_ap}\mathcal{L}_D(X_D,\theta)&=\sum_d m_d\langle[\lambda_{d,i}r_d(\textbf{x}_i,\theta)]^q\rangle_i\text{, where } \textbf{x}_i\in\Omega_D
\end{align}
\noindent where $\langle\cdot\rangle_i$ denotes the mean operator, $q$ is a positive exponent that controls the smoothness of the loss, and $\textbf{x}_i=(t_i, x_i, y_i, z_i)$ are the data points from the subset $X_D$ sampled from the data domain $\Omega_D$. The index $d=\{u, v, w\}$ identifies the specific variables constrained in the loss, where $u, v, w$ represent the velocities in the $x, y, z$ directions. The residual $r_d(\textbf{x}_i, \theta) = |\hat{d}(\textbf{x}_i) - d(\textbf{x}_i, \theta)|$ quantifies the difference between the experimental observation $\hat{d}(\textbf{x}_i)$ and the network prediction $d(\textbf{x}_i, \theta)$ at point $\textbf{x}_i \in \Omega_D$. We use local weights ($\lambda_{d, i}$) to balance the point-wise contribution of the residual $r_d(\textbf{x}_i, \theta)$ and global weights ($m_d$) to scale the averaged value of subcomponent $d$.

The velocity and vorticity boundary conditions are imposed by the boundary loss ($\mathcal{L}_B$), described in Equation~\ref{bcs_eq}.
\begin{align}
\label{bcs_eq}
\mathcal{L}_B(X_B,\theta)&=\sum_b m_b\langle[\lambda_{b,j}r_b(x_j,\theta)]^q\rangle_j\text{, where } x_j\in\Omega_B
\end{align}
\noindent here, $\Omega_B = \{(t, x, y, z) \in \Omega; y = 0 \text{ or } y = 1\}$ is the boundary domain corresponding to the top and bottom walls of the domain $\Omega$. To impose the no-slip boundary conditions and no-normal vorticity, we set $b = \{u, v, w,\omega_y\}$, with residuals $r_b(x_j, \theta) = |b(x_j, \theta)|$, local weights $\lambda_{b, j}$, and global weights $m_b$. 

To enforce our governing equations, we rewrite equations~\ref{Mom_eq_vw},~\ref{cont_vw},~\ref{Temp_Eq} and ~\ref{Eq_id_w} in their residual form and impose them iteratively by minimizing the loss function described in equation~\ref{PDE_loss}. In the following section, we follow~\cite{jin2021nsfnets} and use subscript notation to define the derivative with respect to each component (e.g., $u_z=\frac{\partial u}{\partial z}$). The residuals for the momentum and temperature residual equations are defined as:
\begin{align}
\footnotesize
r_{Mx}&=a_t+ua_x+va_y+wa_z-(au_x+bu_y+cu_z+\sqrt{\frac{Pr}{Ra}}(a_{xx}+a_{yy}+a_{zz})-T_z)\\
r_{My}&=b_t+ub_x+vb_y+wb_z-(av_x+bv_y+cv_z+\sqrt{\frac{Pr}{Ra}}(b_{xx}+b_{yy}+b_{zz}))\\
r_{Mz}&=c_t+uc_x+vc_y+wc_z-(aw_x+bw_y+cw_z+\sqrt{\frac{Pr}{Ra}}(c_{xx}+c_{yy}+c_{zz})+T_x)\\
r_{T}&=T_t+uT_x+vT_y+wT_z-\sqrt{\frac{1}{PrRa}}(T_{xx}+T_{yy}+T_{zz})\\
\end{align}

\noindent where $a=w_y-v_z$, $b=u_z-w_x$ and $c=v_x-u_y$ are the vorticity components in the $x,y$ and $z$ directions. Similarly, we rewrite the conservation of mass and the vector identity (equation ~\ref{Eq_id_w}) into their three components as follows:

\begin{align*}
r_{DFu}&=u_x+v_y+w_z\\
r_{VIx}&=c_y-b_z+(u_{xx}+u_{yy}+u_{zz})\\
r_{VIy}&=a_z-c_x+(v_{xx}+v_{yy}+v_{zz})\\
r_{VIz}&=b_x-a_y+(w_{xx}+w_{yy}+w_{zz})\\
\end{align*}

We enforce the physical knowledge by minimizing the residual equations using a combined loss function:

\begin{align}
\label{PDE_loss}
\mathcal{L}_E(X_E,\theta)&=\sum_e m_e\langle[\lambda_{e,l}r_e(x_l,\theta)]^q\rangle_l\text{, where } x_l\in\Omega
\end{align}

\noindent here, $e = \{M_x, M_y, M_z,DF_u, T, VI_x, VI_y, VI_z,\}$ identifies the residuals from the momentum $M_\beta$, continuity $DF$, temperature $T$, and the vector identities $VI$ in the $\beta=\{x,y,z\}$ directions. We define the residual for subcomponent $e$ and point $x_l \in \Omega$ as $r_e(x_l, \theta) = |e(x_l, \theta)|$, and balance its point-wise and averaged contribution to $\mathcal{L}_E$ using local multipliers $\lambda_{e, l}$ and global weights $m_e$, respectively. The remaining constraint (i.e., equation~\ref{cont_vw}) is strictly satisfied by the model's definition.

\subsubsection{Sequential Training} Training AIVT involves minimizing 15 objective functions, respectively, which significantly complicates the optimization process.
To simplify this problem, we propose a sequential learning approach that divides the training into four stages (See Figure~\ref{all_results}(C)). To ensure a smooth transition, we reinitialize the optimizer and local multipliers at the beginning of each phase. 

In the first stage, we solve a purely data-driven problem where the models only learn the data and the boundary conditions. In particular, for the first ($1/6$) of the training iterations, we minimize the following loss function:
\begin{align}
 \mathcal{L}=\mathcal{L}_D+\mathcal{L}_B+\mathcal{L}_T^*
\end{align}

Where the data loss $\mathcal{L}_D$ and the boundary loss $\mathcal{L}_B$ are described in equations~\ref{data_eq} and ~\ref{bcs_eq} with $q=2$ (i.e., Mean Squared error). $\mathcal{L}_T^*$ is a constraint that guides the model during the initial iterations. In particular, this loss guides the predicted temperature in the core region ($0.1<y<0.9$)to match the theoretical average $\bar{T}=0.5$.  This constraint is imposed softly since it's scaled with a small global weight $m_T^*=m_d/100$ that decays as we approach the final iterations. Our results have shown that using $\mathcal{L}_T^*$ is not necessary to reproduce our results; however, it helps stabilize the model for different initialization.

In the second step ($1/6-2/6$), we partially include the equation constraints as:
\begin{align}
 \mathcal{L}=\mathcal{L}_D+\mathcal{L}_B+\mathcal{L}_T^*+\mathcal{L}_E^*
\end{align}
$\mathcal{L}_E^*$ is a loss function that enforces "partial physics." $\mathcal{L}_E^*$ shares equation~\ref{PDE_loss} formulation but differs from it since it enforces the PDE equations using a lower Rayleigh number $Ra=Ra/100$. This technique enables us to learn and capture the diffusive features of the flow. The first two stages can be considered an initialization or ``warm-up" stage, where the model learns a similar function, facilitating and enabling convergence to the actual solution.

In the third stage, $(2/6-5/6)$ of the training iterations, we use the final Rayleigh number and learn the turbulent flow by minimizing the full loss function:
\begin{align}
 \mathcal{L}=\mathcal{L}_D+\mathcal{L}_B+\mathcal{L}_E
\end{align}
\noindent where $\mathcal{L}_E$ is described in equation~\ref{PDE_loss}. In the last stage (i.e., last quarter), we set $q=1$, which switches the MSE to the mean absolute error MAE, helping refine the details of the learned solution.

\subsection{Residual-Based Attention with resampling (RBA-R)} One of the main challenges in training neural networks is that the residuals (i.e., point-wise errors) may get overlooked when calculating a cumulative loss function \cite{anagnostopoulos2024residual,mcclenny2023self}. To address this issue, research has proposed using direct and indirect methods. Direct methods employ local weights($\lambda_{\alpha,i}$) to balance specific residuals' point-wise contribution within each loss term $\alpha$. On the other hand, indirect methods rely on resampling or refining the high-error regions \cite{wu2023comprehensive}. In this study, we extend residual-based attention (RBA) weights to scale the point-wise residual (i.e., direct method) and to resample the high error regions (i.e., direct method). The update rule for an RBA weight ($\lambda_{\alpha, i}$) for the loss term $\alpha$ and point $x_i$ is based on the exponentially weighted moving average of the residuals defined as:

 \begin{equation}
 \lambda_{\alpha,i}^{(k+1)} \leftarrow (1-\eta)\lambda_{\alpha,i}^{(k)}+\eta {\frac{r_{\alpha,i}^{(k)}}{\;\,\,\lVert \bm{r}_\alpha^{(k)} \rVert_{\infty}}}, \ \ i \in \{0, 1, ..., N\},
 \label{Update_RBA_ap}
 \end{equation}
   
  \noindent where $k$ is the iteration, $N$ is the number of training points, $r_{\alpha,i}$, is the residual of loss term $\alpha$ for point $i$, $\eta$ is a learning rate, and $\gamma$ is a decay term that reduces the contribution of the previous iterations. This formulation induces RBA to work as an attention mask that helps the optimizer focus on capturing the spatial or temporal characteristics of the specific problem \cite{anagnostopoulos2024residual,anagnostopoulos2024learning}. 

   Generally, obtaining an attention mask requires processing as many multipliers as training points, which may hinder its optimal application for large datasets due to its computational cost. To address this issue, we propose using the obtained multipliers to resample the critical points. As shown in equation~\ref{Update_RBA_ap}, RBA weights contain historical information about the high-error regions, making them suitable for defining a probability density function $p_{\alpha}(\textbf{x})$. Building on the previous studies \cite{lu2021deepxde,wu2023comprehensive}, we define $p(\textbf{x})^{(k)}$ at iteration $k$ as follows:
   \begin{equation}
   p_{\alpha}^{(k+1)}(\textbf{x})=\frac{(\bm{\lambda}_{\alpha}^{(k)})^{\nu}}{\mathbb{E}[(\bm{\lambda}_{\alpha}^{(k)})^{\nu}]}+c
 \label{Update_pdf_ap}
 \end{equation} 
 \noindent where $(\bm{\lambda}_{\alpha}^{(k)})^{\nu}=\{\lambda_0^{(k)\nu},\lambda_{\alpha,1}^{(k)\nu},...,\lambda_{\alpha,N}^{(k)\nu}\}$ are the RBA weights of loss term $\alpha$. The exponent $\nu$ is an integer that controls the standard deviation of $p^k(\textbf{x})$, and $c>0$ is a scalar that ensures that all points are eventually resampled.

\section{Implementation details and hyperparameter settings} The cKAN used in this paper was composed of 7 hidden layers with 64 neurons per layer using a polynomial order $k=5$. The total number of training points for data, boundary conditions, and collocation points (i.e., to evaluate or PDE) was $3.8\times10^{5}$, $10^{5}$, and $8.1  \times10^{5}$, respectively. The AIVT model was trained with $3\times10^5$ mini-batch training iterations divided into four stages. The batch size is 5000 data points, and the specific points are resampled at every training iteration using an RBA-based PDF as described in the methods section. During the data-driven stage (first 1/6) iterations, we used the AdamW optimizer, and in the remaining stages, using the Adam optimizer. We applied an inverse-time decay learning rate that starts on $lr_0=3\times10^{-4}$ and ends at $lr_f=10^{-5}$ with a decay rate of $0.9$. In the loss function, we used global weights $m_{\alpha}$ to balance the contribution of each loss subterm. We modify these values during training, enabling us to enforce the proposed sequential training approach easily. The specific global weights are chosen so that the relative contribution of all loss terms is in the same order of magnitude. The specific values are described in Table~\ref{global}. 

Additionally, we use RBA as a local weight to balance the contribution of each residual point. For RBA, we use a decay rate $\gamma=0.999$ and a learning rate $lr=0.1$, which induces an upper bound of one hundred. Since RBA contains the historical information of the residuals, we use them to define a PDE used to resample the training point at each iteration; for this step, we use $\nu=2$ and $c=0.5$. All our results were generated using a single GPU NVIDIA A100-SXM-80GB using JAX as the machine learning framework.

For our ablation study, we fix the seed, which ensures that all models' parameters are initialized identically. For the MLP, we use $sin$ as an activation function and choose an architecture that approximately matches the total number of parameters of our cKAN network, namely, seven layers and 143 neurons per layer. Using this architecture, the number of parameters for cKAN and MLP are 143360 and 143143, respectively. 

\begin{table}[H]
\centering
\caption{Global Weights}

\begin{tabular}{lccc}
 Stage&Loss Term & Sub-term & Global Weight ($m_{\alpha}$)\\
\midrule
1 &Data& $u$& 1000\\
1 &Data& $v,w$ & 10000\\
1 &Data& $T^*$ &10\\
1 &Boundaries& $u,v,w,w_y$& 10\\
1 &Equations& Momentum $x,z$& $10^{-10}$\\
1 &Equations& Momentum $y$& $10^{-10}$\\
1 &Equations& Temperature $T$& $10^{-10}$\\
1 &Equations& Conservation of mass& 1\\
1 &Equations& Vector Identities $x,y,z$ &1\\
\hline
2 &Data& $u$& 100\\
2 &Data& $v,w$ & 1000\\
2 &Data& $T^*$ &0.1\\
2 &Boundaries& $u,v,w,w_y$& 10\\
2 &Equations& Momentum $x,z$& 1\\
2 &Equations& Momentum $y$& 10\\
2 &Equations& Temperature $T$& 10\\
2 &Equations& Conservation of mass& 1\\
2 &Equations&Vector Identities $x,y,z$ &1\\
\hline
3 &Data& $u$& 10\\
3 &Data& $v,w$ & 100\\
3 &Data& $T^*$ &0.01\\
3 &Boundaries& $u,v,w,w_y$& 10\\
3 &Equations& Momentum $x,z$& 1\\
3 &Equations& Momentum $y$& 10\\
3 &Equations& Temperature $T$& 10\\
3 &Equations& Conservation of mass& 1\\
3 &Equations&Vector Identities $x,y,z$ &1\\
\hline
4 &Data& $u$& 1\\
4 &Data& $v,w$ & 10\\
4 &Data& $T^*$ &0.001\\
4 &Boundaries& $u,v,w,w_y$& 1\\
4 &Equations& Momentum $x,z$& 0.1\\
4 &Equations& Momentum $y$& 1\\
4 &Equations& Temperature $T$& 1\\
4 &Equations& Conservation of mass& 0.1\\
4 &Equations&Vector Identities $x,y,z$ &0.1\\
\bottomrule
\end{tabular}
\label{global}
\end{table}

\newpage
 \bibliographystyle{elsarticle-num} 
 \bibliography{cas-refs}





% \end{thebibliography}
\end{document}